\documentclass[final,1p,times]{elsarticle}

\usepackage{hyperref}
\usepackage{subfigure}
\usepackage{tikz}

\newcommand{\avg}[1]{\langle #1 \rangle}

\journal{Information Sciences}

\begin{document}

\begin{frontmatter}

\title{Network Structure and Resilience of Mafia Syndicates}

\author[mymainaddress]{Santa Agreste}
\author[mymainaddress,mysecondaryaddress]{Salvatore Catanese}
\author[thirdaddress]{Pasquale De Meo}
\author[forthaddress]{Emilio Ferrara\corref{mycorrespondingauthor}}
\ead{ferrarae@isi.edu}
\author[mymainaddress]{Giacomo Fiumara}

\cortext[mycorrespondingauthor]{Corresponding author}

\address[mymainaddress]{Dept. of Mathematics and Informatics, University of Messina, Italy.}
\address[mysecondaryaddress]{Dept. of Mathematics and Informatics, University of Catania, Italy.}
\address[thirdaddress]{Dept. of Ancient and Modern Civilizations, University of Messina, Italy.}
\address[forthaddress]{Information Sciences Institute, University of Southern California,  Marina del Rey, CA, USA.}

\begin{abstract}
In this paper we present the results of the study of Sicilian Mafia organization by using Social Network Analysis. 
The study investigates the network structure of a Mafia organization, describing its evolution and highlighting its plasticity to interventions targeting membership and its resilience to disruption caused by police operations. 
We analyze two different datasets about Mafia gangs built by examining different digital trails and judicial documents spanning a period of ten years: the former dataset includes the phone contacts among suspected individuals, the latter is constituted by the relationships among individuals actively involved in various criminal offenses. 
Our report illustrates the limits of traditional investigation methods like tapping: criminals high up in the organization hierarchy do not occupy the most central positions in the criminal network, and oftentimes do not appear in the reconstructed criminal network at all.
However, we also suggest possible strategies of intervention, as we show that although criminal networks (i.e., the network encoding mobsters and crime relationships) are extremely resilient to different kind of attacks, contact networks (i.e., the network reporting suspects and reciprocated phone calls) are much more vulnerable and their analysis can yield extremely valuable insights.
\end{abstract}


\end{frontmatter}


\section{Introduction}
\label{sec:intro}
Sicilian Mafia (often known as {\em Cosa Nostra}) is a criminal organization which originated in Sicily and, after decades of emigration waves, it is now spread worldwide \cite{sonnino1877sicilia, mastrobuoni2012organized, mcgloin2005policy}.

Police investigations revealed that Mafia is a loose confederation of smaller syndicates (called  ``cosche'', ``clan'' or ``Families'') such that each syndicate takes the control of a specific territory (usually a town or a part of it) by organizing and overseeing illegal activities. Members of a Mafia syndicate can be both mobsters and associates (i.e., people like drug-dealers, murders or corrupted politicians who are not part of the syndicate but contribute to its illicit activities). Mafia syndicates show a strong hierarchical organization \cite{mastrobuoni2012organized}: on the top of the organization we have a ``boss'', who is aided by an ``underboss'' and by various ``liutenants'' who head branches of the Mafia syndicate. The boss also commands a crew of ``soldiers'' (often known as {\em picciotti}) who commit acts of violence like intimidation, threats and murders.

Due to its normative structure as well as strong ties with finance, entrepreneurs and politicians, Mafia has now risen to prominence as a worldwide criminal organization by controlling many illegal activities like the trade of cocaine, money laundering or illegal military weapon trafficking \cite{cayli2013italian}.


Understanding the structure of Mafia syndicates, unveiling the functional role of each of their members and quantify the ability of a Mafia syndicate to react to the detention of its members are crucial steps to effectively fight and dismantle Mafia syndicates.
In the latest years, various researchers \cite{mastrobuoni2012organized, FerraraMCF14, morselli2003career} illustrated the benefits of using Social Network Analysis \cite{Newman10} to study the structure criminal organizations.

The adoption of methods from Social Network Analysis in the study of criminal organizations has strong theoretical and practical motivations: studies from sociological literature (known as {\em social facilitation models} \cite{mastrobuoni2012organized}) point out that the membership of an individual to a crime gang enormously amplifies her/his tendency to criminal behaviors \cite{thornberry1993role}.
Destroying the network structure associated with a criminal organization is central to prevent individuals from committing crimes and lower delinquency rates.


The first step to analyze Mafia syndicates by means of Social Network Analysis tools is to collect a sufficiently large data sample describing the various units composing the syndicate and their operations.
Interactions among mobsters materialize under various forms: for instance two mobsters can be tied if they committed together the same crime or if they have been seen together in the same sighting.
A powerful and well-known investigation method is {\em tapping}, i.e., the procedure of recording information flow among suspected criminals which has been sent using any type of electronic media, like phone calls (from both fixed lines and mobiles), emails, SMS messages and private communications over Social Media platforms. Tapping has proven to be effective for preventing and solving many crimes like terrorism, drugs, kidnapping and political corruption.

Tapping has been extensively used also in case of Mafia related investigations, but, if used alone, it may fail to reliably capture the structure of a Mafia syndicate:  newspapers, for instance, report that the boss of Mafia syndicates often reveal their whereabouts to just few gang members and, in many cases, they issue orders and communications through handwritten notes known as {\em pizzini}. \footnote{See \url{http://news.bbc.co.uk/2/hi/europe/4899512.stm}}

A promising investigation strategy requires to supplement information collected by tapping with data generated by other methods of investigation like video surveillance, use of informants and under-cover agents, interview to subjects, analysis of bank transactions and so on.
By gluing together this piece of information, we can capture a more detailed picture of the structure of a Mafia syndicate.
Unfortunately, the information cited above is the outcome of a long, expensive and often dangerous investigation process which often spans years, or, in certain cases, even decades.


After examining several types of judicial documents spanning a ten years period (like judgments, verdicts, depositions and inquisitions, and so on) we built two datasets about Mafia gangs operating in the North of Sicily.

We started by collecting phone calls among suspected individuals and this allowed us to build a network called {\em contact network} $N_{\mathrm{con}}$  in which each individual was associated with a vertex and an edge between two vertices denotes the existence of at least one reciprocated phone call between the individuals associated with that vertices. The network $N_{\mathrm{con}}$ contained 1, 716 vertices and 8,481 edges.

By means of further investigation, we were able to identify {\em crime relationships}: we say that a crime relationship exists between two individuals if they took part in the same criminal offense or if they have been seen together during a sighting. Criminal relationships were then mapped on a second network called {\em criminal network} $N_{\mathrm{cri}}$. The network $N_{\mathrm{cri}}$ contained only 104 vertices and 2,596 edges:  all but 6 individuals in the criminal network were also present in the contact network. This means that the original dataset contained almost all mobsters but there were mobsters who were part of the Mafia syndicate but who never used mobiles or fixed lines to communicate.

The availability of these datasets offered us the unprecedented opportunity of understanding the actual structure of a Mafia syndicate and to quantify how well it is able to react to police operations leading to the detention of some of its members.
In a first stage of our research, we studied the structural properties of $N_{\mathrm{cri}}$ and $N_{\mathrm{con}}$. Our first goal was to
understand whether meaningful differences arise between the structural features of the two networks.

Subsequently, we investigated and compared the robustness of $N_{\mathrm{con}}$ and $N_{\mathrm{cri}}$. We simulated a police operation leading to the arrest of a fraction $f$ of individuals from the two networks and we studied how these perturbations impacted on the structure of both $N_{\mathrm{con}}$ and $N_{\mathrm{cri}}$.
Individuals were selected either randomly or on the basis of their centrality in the network. To this end, we used three different centrality metrics, namely Degree Centrality ($DC$), Betwenness Centrality ($BC$) and Closeness Centrality ($CC$).
We considered two type of operations, namely: {\em (i) Parallel Attack}, i.e., we assumed that a fraction $f$ of individuals were {\em simultaneously} deleted from the network along with their connections and {\em (ii) Sequential Attack}. i.e., we supposed to iteratively neutralize individuals along with their connections from the network until a fraction $f$ of individuals has been neutralized.
To measure the effectiveness of each operation, we computed two parameters, namely the size of the Strongly Connected Component ($\mathtt{SCC}$) of each network and the Average Path Length ($\mathtt{APL}$) (defined as the mean of shortest path lengths in the network).

The main findings of our analysis can be summarized as follows:

\begin{enumerate}

\item We found that 98 (out of 104) members of  $N_{\mathrm{con}}$ were also members of $N_{\mathrm{cri}}$; there were also 6 mobsters who appeared in 
$N_{\mathrm{cri}}$ but were not recorded in $N_{\mathrm{con}}$. This proves the presence in the contact network of few criminals who do not use phones to communicate because they consider phone calls unreliable and unsafe.

\item The degree distribution in $N_{\mathrm{con}}$ followed a power law $k^{-\alpha}$ with $\alpha = 2.5$. By contrast, the degree distribution in $N_{\mathrm{cri}}$ was almost uniform and about $76.92\%$ of $N_{\mathrm{cri}}$ affiliates had a degree ranging between 15 and 85. With the help of police officers, we observed that leaders in the Mafia syndicate were the individuals in $N_{\mathrm{cri}}$ showing the lowest degree.
Therefore, the top elements in a Mafia syndicate do not occupy the most central positions in the criminal network.

\item Social relationships were dense both in $N_{\mathrm{con}}$ and $N_{\mathrm{cri}}$. To this end, we computed the Average Clustering Coefficient of each vertex as function of its degree and we found that it was always bigger than 0.6 (i.e., more than 5 times the value measured on social networks like Facebook \cite{ugander2011anatomy, catanese2011crawling}). This is likely to depend on the recruiting policies of Mafia syndicates which impose the existence of intermediaries to enable an individual to join a Mafia syndicate.

\item In case of a parallel police operation, we observe that targeted attacks are able to quickly destroy the strongly connected component of $N_{\mathrm{con}}$. In particular, $DC$ has the most disruptive effect on $\mathtt{SCC}$. 		
In contrast, $N_{\mathrm{cri}}$ showed an exceptional degree of robustness  independently of the adopted centrality index.

\item In case of sequential police operations, the $CC$ has the most disruptive effect on $\mathtt{SCC}$ and this happens both in $N_{\mathrm{con}}$ and in $N_{\mathrm{cri}}$. $CC$ is still the best option if the goal is to increase $\mathtt{APL}$ in $N_{\mathrm{con}}$; by contrast, $DC$ yields the largest increase in $\mathtt{APL}$ if applied on $N_{\mathrm{cri}}$.

\end{enumerate}
The plan of the paper is as follows: in Section \ref{sec:related} we discuss the related literature. In Section \ref{sec:background} we provide some basic definitions which will be largely used throughout the paper. In Section \ref{sec:structure} we describe the main structural features of both the Contact and the Criminal Network while Sections \ref{sec:resilience-parallel} and \ref{sec:resilience-sequential} are devoted to investigate the resilience of the Contact and Criminal networks under parallel and sequential attacks, respectively. Finally, in Section \ref{sec:conclusions} we draw our conclusions and illustrate our future research plans.

\section{Related Literature}
\label{sec:related}

In this section we review the scientific literature related to our approach. 
We start discussing how Social Network Analysis techniques have been applied to study Mafia-related organizations (Section \ref{sub:mafia-social}). We then illustrate approaches concentrating on how the power of a criminal organization depends on the social relationships among its members (Section \ref{sub:social-interactions-crime}). One of the major contributions of our study, in fact, consists of exploring the effects associated with the dismissal of one (or more) members of a criminal gang.    

\subsection{Social Network Analysis and Mafia syndicates}
\label{sub:mafia-social}

One of the early reports on the structure of Mafia syndicates dates back to 1876 and is due to the Italian deputy Leopoldo Franchetti \cite{sonnino1877sicilia}, who depicted the Mafia as a criminal organization deeply rooted in Sicilian society. Franchetti argued  that Mafia was impossible to destroy unless a deep change in Sicilian social institutions would occur.

Such a study has deeply influenced prosecuting magistrates, politicians, criminologists and sociologists committed to fighting Mafia.
Mafia syndicates are organized according to rigid normative structures, being perhaps the {\em Mafia Decalogue} the most popular code of conduct.
According to that Decalogue, mobsters must respect each other: for instance, it is forbidden to appropriate money if it belongs to other members of the same syndicate or to other families.
Ties among mobsters belonging to the same syndicate are very strong: in some cases they are related by blood and, in any case, the gang comes before their birth family.

Because of the rich and strong web of relationships
among mobsters, the analysis of the social structure of a Mafia syndicate is of great scientific interest and it well explains why  Social Network Analysis methods have been extensively applied to the study of Mafia syndicates.

For instance, Morselli \cite{morselli2003career} studied the connections within a New York-based family (the Gambino family). The study focused on the career of one of its members, Saul Gravano. One of the main Morselli's findings  is that Gravano's ability of building and extending over time his personal network of contacts was a key factor to climbing the Gambino's family organization.
Natarajan \cite{natarajan2006understanding} studied a dataset consisting of 2,408 wiretap conversations gathered during the prosecution of a heroin-dealing Mafia syndicate in New York. Starting from available data, the author built a network of phone calls, which revealed a group of 294 individuals forming the core of the criminal organization. Natarajan showed that most of the group members had very limited contacts with others in the group.

Other relevant studies are reported by Sarnecki \cite{sarnecki2001delinquent} (who applied Social Network  Analysis to study co-offending behaviors among Swedish teenagers) and by McGloin \cite{mcgloin2005policy} (who analyzed the network structure of street gangs in Newark, New Jersey).

Social Network Analysis is not only a tool to describe the structure and functioning of a criminal organizations but it has been largely employed in the construction of crime prevention systems \cite{chen2004crime}. 
For instance, Xu and Chen \cite{xu2005crimenet} jointly applied Social Network Analysis with hierarchical clustering algorithms; the proposed approach worked in two stages: first, a criminal network was partitioned into subgroups by means of a clustering algorithm. Then, block modeling techniques have been used to extract interaction patterns between these subgroups.
A further application of Social Network Analysis to crime detection is reported in \cite{drezewski2015application}, which focused on  money laundering.

Social Network Analysis tools were finally employed to identify leaders within a criminal organization.
For instance, Mastrobuoni and Patacchini \cite{mastrobuoni2012organized} used a dataset recording criminal profiles of 800  Mafia members active in the United States from 1950s to 1960s to investigate the structure of criminal ties between mobsters.
Various features were considered like family relationships, legal and illegal activities, to predict the criminal rank of a mobster.

In our previous work we focused on the joint application of Social Network Analysis tools and advanced Data Visualization techniques \cite{FerraraMCF14, catanese2013forensic}. We described a software system which was able to extract criminal organizations from a network recording mobile phone calls and we
combined statistical network analysis primitives, community detection algorithms, and visual exploration tools, to unveil the structure of criminal networks hidden in communication data.

This paper introduces many novelties with respect to the approaches cited above. In fact, our analysis focuses on two datasets which roughly cover the same time interval and refer to the same geographical area. The first dataset records phone calls while the second one is about crime relationships and, as will become clear in the following, the networks extracted from each dataset show deep differences from a structural viewpoint.
Our work highlights the limits of tapping as investigation method to fight against Mafia gangs and it shows that the most prominent criminals do not occupy the most central positions in the criminal network.
The procedure we followed to build the datasets in this paper essentially relies on the analysis of judicial documents like  verdicts of depositions. An approach to collecting crime related data which is ortoghonal to ours is described by Furtado et al. \cite{furtado2010collective}.
In that paper, the authors describe {\em WikiCrime},  a Web  application that enables its users to directly report crimes on a specific geographical area and temporal window, or to search for a specific crime occurred in the past. One of the core features of WikiCrime is its ability to give more transparency and diffusion to criminal information and to prevent crimes. As claimed by the authors, WikiCrime is also effective
to reduce {\em under reporting}, i.e., the fact that some crimes are not notified to law enforcement authorities.
WikiCrime integrates a reputation module to verify the credibility of generated information: on the one hand, in fact, the collaboration of large masses of users enables to quickly and cheaply collect vast amount of data. On the other hand, the source of available data is often unknown and, therefore, it is hard to determine if the information is credible and accurate.

Concluding, we point the interested reader to the informative and comprehensive review recently compiled by D'Orsogna and Perc \cite{dorsogna2015statistical} that summarizes current efforts in computational modeling of crime from a statistical physics perspective.

\subsection{The power of criminal organizations and social interactions}
\label{sub:social-interactions-crime}

Many researchers studied what are the best policies to fight (and hopefully dismantle) a criminal organization.
Most of these studies highlight the importance of social relationships as a multiplier of the aptitude of single individuals to commit crimes.
For instance, one of the early contribution is due to Sah \cite{sah91}, who proposed a crime model based on social interactions. The key point of that study  is that the severity of punishment perceived by an individual as a consequence of her/his illicit behaviors depends on her/his social setting. 
As a result some individuals (under the influence of their peers, the social environment they live in and the institutions with whom their interact) may consider the punishment as not severe and this is more conducive to criminal actions.
An interesting study by Gleaser et al. \cite{glaeser1996crime} classified individuals in a criminal network as {\em conformist} (if they simply imitate the behaviors of their peers) and {\em non-conformist} (if they decide on their own to commit/not commit crimes).
The studies reported above highlight that the structure of social ties among members of the same community as well as the culture individuals have been exposed to may have a crucial impact on their tendency to commit crimes. The main question deriving from these studies is how to perturb a criminal network to reduce the aptitude of its members to commit crime. 

Ballester and collaborators \cite{ballester2006s} suggested a {\em key-player} policy which targets at removing the criminal who reduces the most the level of criminality in a gang. Such a policy was more effective than traditional punishment policies.  Borgatti \cite{borgatti2006keyplayers} defined a different approach for key-players finding  based on qualitative features of vertices rather than a mere quantitative evaluation of their centralities. Unfortunately, such an approach requires access to further information that might not be readily available to the investigators, or might be dangerous to collect in the context of criminal investigation.

Liu et al. \cite{liu2012criminal} analyzed delinquent networks of adolescents in the United States with the goal of detecting the criminal(s) who, once removed, generate the highest possible reduction in aggregate crime level. 
They found that in delinquent adolescent networks, key players are more likely to be male, have less educated parents, are less attached to religion and feel socially more excluded. 

In this paper we consider a similar problem, i.e., we focused on finding what police strategy has the most disruptive effect on the structure of a Mafia gang. Our analysis and the collaboration with experts from local law enforcement agencies highlight that, in Mafia syndicates, the key players are not the best connected mobsters: in fact, the criminals occupying leadership roles in Mafia gangs often prefer not to use phones to communicate. In addition, we observed that bosses were not concentrated on a specific region of the criminal network but they were uniformly spread
in the network. This encodes the fact that bosses often belong to different family units. As a consequence, the task of arresting bosses is extremely difficult and dangerous. As a further result, this paper shows that the criminal network (i.e., the network encoding mobsters and crime relationships) is extremely resilient to different kind of attacks while the contact network (i.e., the network recording suspected individuals and reciprocated phone calls) is much more vulnerable.

\section{Background}
\label{sec:background}

In this section we briefly introduce centrality scores (Section \ref{sub:centrality-scores}), and then we illustrate the concept of network robustness (Section \ref{sub:network-robusteness}).

In the following we define a network $N = \langle V, E\rangle$ as a pair in which $V$ is the set of vertices and $E$ is the set of edges.
The symbol $\langle i, j \rangle$ denotes an edge in $E$.

A network $N$ can be represented through its {\em adjacency matrix} $\mathbf{A}$, which is defined as follows: $\mathbf{A}_{ij} = 1$ if (and only if) there is an edge going from the vertex $i$ to the vertex $j$, $0$ otherwise. In the following we suppose that $\mathbf{A}$ is {\em symmetric}, i.e., if an edge $\langle i, j  \rangle$ belongs to $E$, then the edge $\langle j, i  \rangle$ belongs to $E$ too.

\subsection{Centrality in Networks}
\label{sub:centrality-scores}

The centrality of an individual, represented by a vertex \textit{i,} in a network \textit{N} is a measure of the importance of \textit{i} in \textit{N.} A large number of centrality indices has been considered in the literature (see \cite{Newman10} for an excellent review). In our study we focused on three centrality indices, namely: {\em (i)} Degree Centrality (hereafter, $DC$), {\em (ii)} Betwenness Centrality ($BC$) and {\em (iii)} Closeness Centrality ($CC$).

We have chosen these indices because they have a clear geometrical interpretation and, then, the notion of importance they implement is easy to understand. These indices rely on complementary philosophies: as for $DC$, in fact, it is based only on the local connectivity of a vertex and it requires to know only the number of neighbors of a vertex.
More formally, given a vertex \textit{i}, its {\em degree centrality} $DC(i)$ is defined as the number of edges incident onto $i$.
The $BC$ and $CC$ indices are based on the concept of {\em shortest path} (also known as {\em geodesic path}) in a network: given an unweighted and undirected network and a pair of vertices $i$ and $j$ the shortest path connecting $i$ and $j$ is the path consisting of the fewest number of edges. According to the literature \cite{freeman1977set}, shortest paths are preferential pathways to convey and spread messages in a broad range of networks like biological  or social networks.

Some authors \cite{newman2005measure, alahakoon2011k, de2012novel, de2013enhancing} argued that the assumption that information travels along geodesic paths may not hold true in real scenarios: for instance, in case of large online social networks like Facebook, users are agnostic about the whole network topology and, therefore, they are not able to find shortest paths and use them to convey messages. In addition, the computation of shortest paths is computationally unfeasible even on moderately large networks.

In case of criminal networks, however, we guess that geodesic paths are to be preferred to randomly generated paths (i.e., random walks). Criminal networks are much smaller than other types of social networks and we can afford to compute geodesic paths. In addition, to ensure secrecy in the transmission of information, shortest paths are to be preferred to longer ones: it is known that criminals systematically try to expose sensitive information to a minimal number of trusted others.

On the basis of these considerations, we claim that the importance of a vertex $i$ depends on the fraction of shortest paths passing through $i$ because this means that $i$ is able to intercept a relevant portion of the information flowing through the network. This intuition naturally leads to introduce the Betweeness Centrality $BC(i)$ of a vertex $i$ which can be formally defined as follows: let $i$, $u$ and $w$ be any three distinct vertices in a network and let $\sigma_{uw}$ be the number of shortest paths from $u$ to $w$; finally, let $\sigma_{uw}(i)$ be the number of the shortest paths from $u$ to $w$ passing through $i$.
We define $BC(i)$ as:

\begin{equation}
\label{eqn:betwenness}
BC(i)= \sum_{i \neq u \neq w \in V}\frac{\sigma_{uw}(i)}{\sigma_{uw}}
\end{equation}

Alternatively, we may classify $i$ as important if its ``distance'' from other vertices in the network is small because this certifies the ability of $i$ to communicate with other vertices and contribute to the information spreading. There are, of course, various possible definitions of distance between network vertices. The simplest one perhaps consists of measuring the distance between two vertices $i$ and $j$ as the length $\mathtt{SP}(i,j)$ of the shortest path connecting them.
Bearing in mind such a notion of distance, we define the Closeness Centrality $CC(i)$ of $i$ as the reciprocal of the sum of all geodesic distances from $i$ to all other vertices in the network \cite{bavelas1948mathematical}:

\begin{equation}
\label{eqn:closeness-def}
CC(i) = \frac{1}{\sum_{u \in V} \mathtt{SP}(u,i)}
\end{equation}

Some experiments devoted to study collaboration in social groups show that individuals perceived as leaders are those users who generally feature high closeness values \cite{bavelas1948mathematical}.

\subsection{Network Robustness}
\label{sub:network-robusteness}

The study of network robustness (i.e., the ability of a network to react to the failure of some of its components \cite{albert2000error,Newman10}) is strongly linked to studies about the reliability of many biological and artificial systems.

A system $S$, in fact, can often be modeled as a network $N = \langle V, E\rangle$ such that each vertex in $V$ identifies one of the components of $S$ while edges describe interactions among components.
The system $S$ is said to be {\em robust} if it can maintain its functions even if some of its components fail or they stop interacting \cite{albert2000error}.

The robustness of $S$ greatly depends on the topological structure of $N$ and on the existence of multiple paths connecting two vertices in $N$.
To gain an understanding, let us refer to a communication network whose devices exchange messages by means of suitable physical links. In an extreme case, suppose that the network presents a star topology: if we would remove the center of the star (along with the edges coming out from it), we would disconnect the whole network.
Another extreme example occurs if we consider a clique: in such a case the removal of an arbitrary vertex would have no impact on the network functioning.
Between these extreme cases, we observe that the malfunctioning of one or more components (or physical links) may not prevent a source component from correctly interacting with a target one because the source component could find alternative paths. This observation legitimates a popular approach to studying network robustness: we study the fragmentation processes taking place in the network by progressively deleting vertices from $N$ along with their connections  \cite{albert2000error,broder2000graph}. Real networks often include a large, strongly connected component (hereafter $\mathtt{SCC}$) retaining most of the network vertices. After deleting some vertices along with their incident links, other vertices could detach from $\mathtt{SCC}$ to form small clusters (or even remain isolated). Because of this fragmentation process, we expect the network to become less and less connected and this implies that the size of the $\mathtt{SCC}$ decreases. Analogously, the network diameter and the Average Path Length $\mathtt{APL}$ (i.e., the mean of pairwise shortest path lengths) should also increase, thus making communication between vertices more difficult.
According to the literature \cite{ugander2011anatomy}, $\mathtt{APL}$
is a more robust parameter than diameter: in fact, the existence of a long shortest path in the network would imply a large network diameter even if vertices are, on average, only few hops away.

Albert and collaborators \cite{albert2000error} focused on the robustness of two classes of networks, namely: {\em (i) homogeneous networks} in which the probability $P(k)$ that an arbitrary vertex has degree $k$ exponentially decays for large values of $k$; and, {\em (ii) heterogeneous networks}, in which $P(k)$ follows a power law distribution.
Examples of homogeneous networks are the Erd\"os-R\'enyi random graph or the small world model by Watts and Strogatz \cite{Newman10}. 
Examples of heterogeneous networks include the Internet~\cite{faloutsos1999power}, the World Wide Web~\cite{barabasi1999emergence}, and in general most (large-scale) social~\cite{castellano2009statistical}, and techno-social systems~\cite{moreno2002epidemic}.
In their experiments, the authors considered both artificial and real networks (i.e., a sample of Web pages and hyperlinks connecting them) and empirically measured the size of the $\mathtt{SCC}$ and the diameter of the network if an increasingly larger fraction $f$ of vertices was removed \cite{albert2000error}.

Two vertex removal strategies were considered: in the former strategy, vertices were randomly selected while in the latter one the most connected vertices were progressively deleted from the network one by one.

In case of homogeneous networks, no substantial variation in network diameter emerged if vertices were selected at random or in a decreasing order of connectivity.

A completely different behavior was observed for heterogeneous networks: the random removal of vertices had  no effect on the diameter while if the most connected vertices were deleted, then the diameter would quickly increase.

An analogous study was later proposed by Broder et al. \cite{broder2000graph}, which took a sample of the World Wide Web graph and removed Web pages on the basis of the number of their outgoing hyperlinks. Conforming to the previous study \cite{albert2000error}, the authors found that it was sufficient to remove Web pages referred by at least other five pages to destroy the Web connectivity \cite{broder2000graph}.

\section{The structure of Contact and Criminal Networks}
\label{sec:structure}

As pointed out in the introduction, we used two datasets (see Table~\ref{tab:CNmeasures}) to analyse the structural properties of Mafia syndicates and their resilience to both random and targeted attacks. The first dataset describes phone calls among suspected criminals located in the North of Sicily. We refined such a dataset by considering further methods of investigation (e.g., stakeout, search in private residence with a search warrant, access to personal bank account information and so on). In this way we built a second dataset recording actual mobsters and crime relationships between them: two mobsters are tied by an edge if they took part in the same criminal offense). Both the two datasets refer to a Mafia syndicate operating in the North of Sicily.

We call Contact Network $N_{\mathrm{con}}=\langle X_{con},E_{con} \rangle$ the network extracted from the first dataset where $X_{con}=\{x_1^{[con]},x_2^{[con]} \ldots, x_n^{[con]} \}$ is the set of individuals (nodes) subject to tapping or found in the phone call logs by law enforcement agencies and $E_{con} \subseteq X_{con} \times X_{con}$ is the set of edges that connect pair of nodes. The set of edges $E_{con}$ is composed of all phone relationships (voice calls, SMS, MMS, etc.). In this work, the edges $E_{con}$ are considered as undirected and unweighted, thus disregarding the orientation and the number of contacts between two any vertices.

Criminal Network $N_{\mathrm{cri}}=\langle X_{cri},E_{cri} \rangle$ is the network obtained from the second dataset where $X_{cri}=\{x_1^{[cri]},x_2^{[cri]} \ldots, x_m^{[cri]} \}$ is the set of individuals subject to deepened investigations by law enforcement agencies by taking into account other kind of relationships among them. 
The set $E_{cri} \subseteq X_{cri} \times X_{cri}$ comprises relationships among components of $N_{\mathrm{cri}}$ not telephone-based such as joint bank transactions, complicity in a crime, and so on.

We call $\mathbf{N}^{[con]}=(n_{ij}^{[con]}) \in \mathbb{N}^{n\times n}$ the {\em adjacency matrix} of $\mathrm{N}^{[con]}$ given by:
\begin{equation}
n^{[con]}_{ij}=\left\{
\begin{tabular}{ll}
1  & if $\langle x^{[con]}_i,x^{[con]}_j\rangle \in E_{[con]},$ \\
0  & \textit{otherwise,}
\end{tabular}
\right.
\end{equation}
and, analogously, with 
$\mathbf{N}^{[cri]}=(n_{ij}^{[cri]}) \in \mathbb{N}^{m\times m}$ the {\em adjacency matrix} of $\mathrm{N}^{[cri]}$ given by:
\begin{equation}
n^{[cri]}_{ij}=\left\{
\begin{tabular}{ll}
1  & if $\langle x^{[cri]}_i,x^{[cri]}_j \rangle\in E_{[cri]},$ \\
0  & \textit{otherwise,}
\end{tabular}
\right.
\end{equation}

Now we define the aggregate network $\mathrm{A}^{[aggr]}=\langle X_{aggr},E_{aggr} \rangle$ where $X_{aggr}=X_{con} \bigcup X_{cri}$, $E_{aggr}=E_{con} \bigcup E_{cri}$ and the aggregated topological {\em adjacency matrix} associated $\mathbf{A}^{[aggr]}=(a_{ij}^{[aggr]}) \in \mathbb{N}^{|X_{aggr}|\times |X_{aggr}|}$ given by:
\begin{equation}
a^{[aggr]}_{ij}=\left\{
\begin{tabular}{ll}
1  & if $n^{[con]}_{ij} \vee n^{[cri]}_{ij} =1,$ \\
0  & \textit{otherwise,}
\end{tabular}
\right.
\end{equation}
of the unweighted network obtained from $N_{\mathrm{con}}$ and $N_{\mathrm{cri}}$  by joining all pairs of nodes i and j which are connected by an edge in at least one network and neglecting the possible existence of multi-ties between a pair of nodes and the nature of each tie as well \cite{battiston2014multiplex}.

\begin{table}[!t]
	\centering
	\begin{tabular}{|l|r|r|r|r|r|r|}
		\hline
		\textbf{Network} & $|V|$ &
		$|E|$ & $\avg{k}$ & $\mathtt{APL}$ & Diameter & $\mathtt{SCC}$\\
		\hline
		Contact Network ($N_{\mathrm{con}}$) & 1716 & 8481 & 9.88 & 2,75 & 6 & 1\\
		Crime Network ($N_{\mathrm{cri}}$) & 104 & 2596 & 49.92 & 1.53 & 3 & 1\\\hline\hline
		Aggregated ($A_{\mathrm{aggr}}$) & 1722 & 11070 & 12.86 & 2.73 & 6 & 1\\
		\hline
	\end{tabular}
	
	\caption{Some statistics about $N_{\mathrm{con}}$ and $N_{\mathrm{cri}}$. For each network we report the number of vertices ($|V|$), the number of edges ($|E|$), the average degree ($\avg{k}$), the Average Path Length ($\mathtt{APL}$), the diameter and the size of the strongly connected component ($\mathtt{SCC}$). We also report the same statistics for the aggregate networks $A_{\mathrm{aggr}}$ obtained from $N_{\mathrm{con}}$ and  $N_{\mathrm{cri}}$ by joining all pairs of nodes $i$ and $j$ which are connected by an edge in at least one network.}\label{tab:CNmeasures}
\end{table}

\begin{figure*}[t!]
	\subfigure[]{
	\label{fig:contact-graphical}
		\begin{minipage}[tb]{0.39\textwidth}
			\centering
			\includegraphics[width=\textwidth]{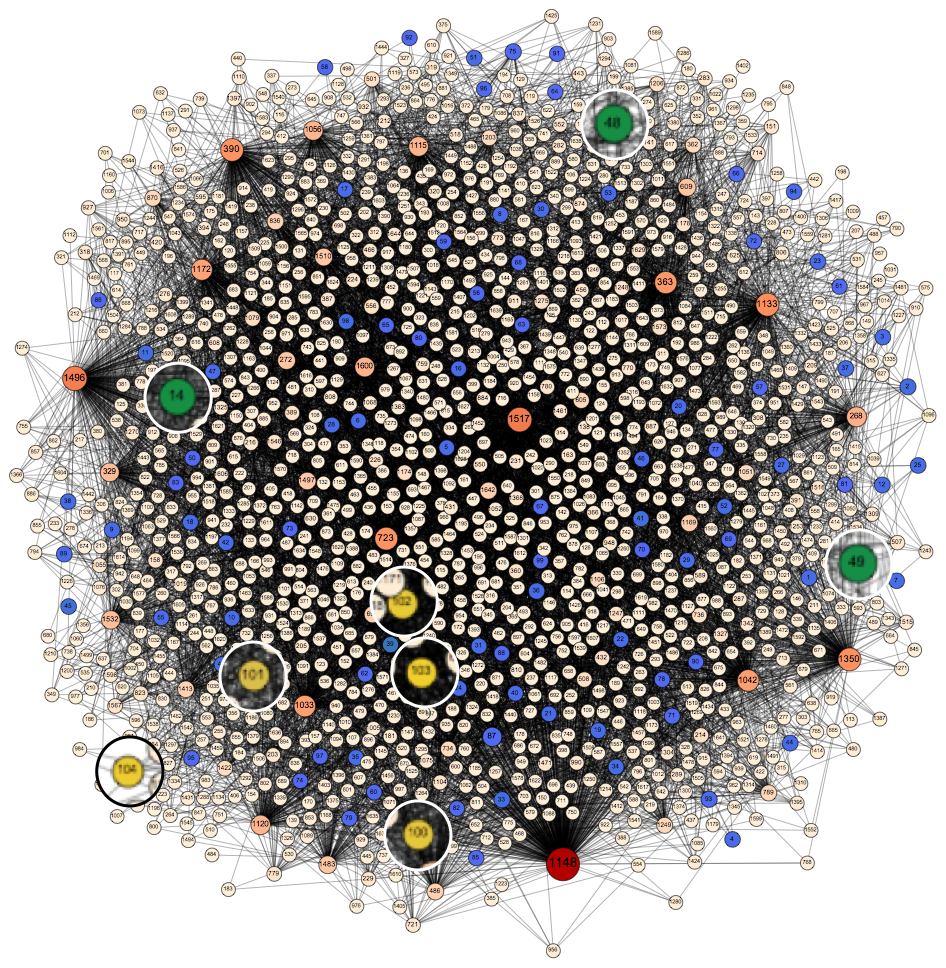}
		\end{minipage}}
	\subfigure[]{
	\label{fig:crimenet-graphical}
		\begin{minipage}[tb]{0.22\textwidth}
			\centering
			\includegraphics[width=\textwidth]{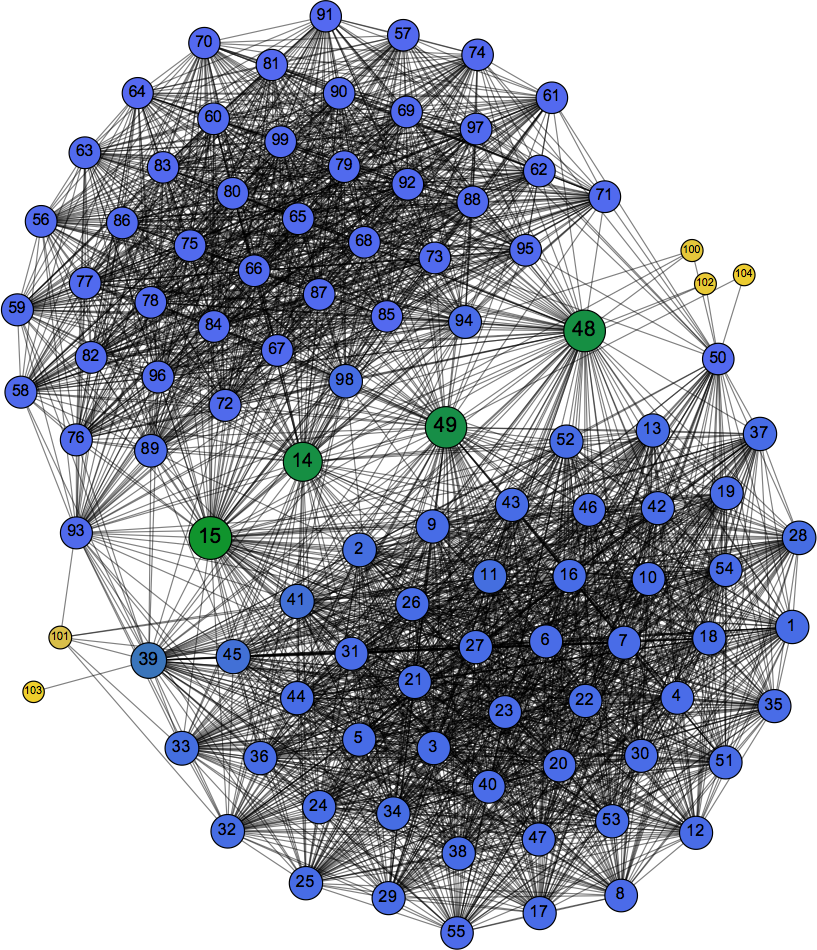}
		\end{minipage}}
	\subfigure[]{
	\label{fig:aggregate-graphical}
		\begin{minipage}[tb]{0.39\textwidth}
			\centering
			\includegraphics[width=\textwidth]{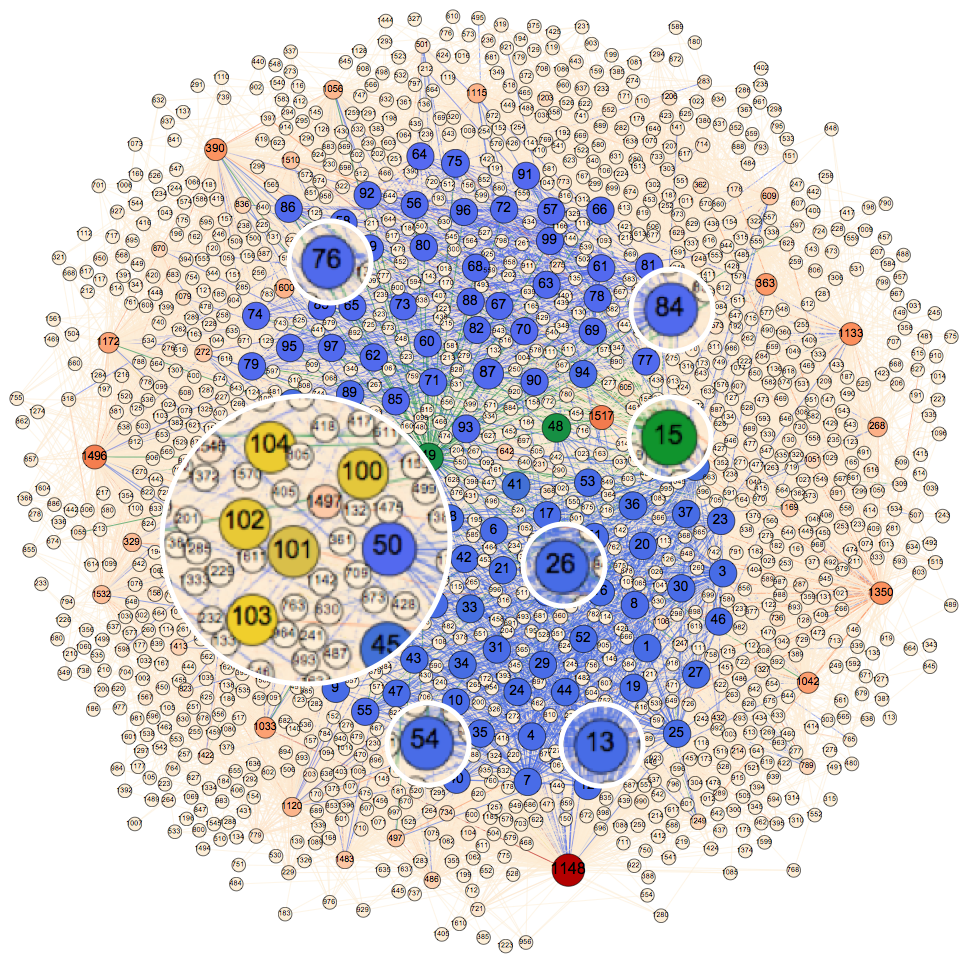}
		\end{minipage}}
\vspace{-0.3cm}
\caption{{\bf Left panel}: We report a graphical representation of $N_{\mathrm{con}}$: here a vertex is associated with a suspected mobster while an edge indicates that the two suspects called each other at least once. Yellow vertices correspond to bosses, green vertices identify lieutenants, and blue vertices identify associates that were later arrested by law enforcement agencies. The size of each vertex is proportional to its degree, and the same holds for the color coding: light yellow is associated to nodes having the minimum degree and red is used for nodes having the maximum degree.   {\bf Center panel}: Graphical representation of $N_{\mathrm{cri}}$, namely mobsters and crime relationship between them (e.g., complicity in a crime, acquaintance, police inspections, bank transactions, etc.) {\bf Right panel}: we show the aggregate network $N_{\mathrm{aggr}}$ where we highlighted vertices corresponding to bosses of the criminal organization (yellow) together with vertices $13,15,26,54,76 \in N_{\mathrm{cri}}$ not belonging to $N_{\mathrm{con}}$, corresponding to mobsters that were never tapped during investigations.}

\label{fig:criminal-contact-network-graphical}
\end{figure*}

In Table \ref{tab:CNmeasures} we report some statistics about $N_{\mathrm{con}}$ and $N_{\mathrm{cri}}$. For each network we indicate the number of vertices ($|V|$) and the number of edges ($|E|$). We observe that $N_{\mathrm{cri}}$ contains only 104 vertices while $N_{\mathrm{con}}$
has 1, 716 vertices. However, $N_{\mathrm{cri}}$ contains 2,596 edges and, therefore, it is much denser than $N_{\mathrm{con}}$ which contains only 8,481 edges. For this reason the average number of edges per vertex is only 9.88 in case of $N_{\mathrm{con}}$ and it amounts to 49.92 in case of $N_{\mathrm{cri}}$. This means that, on average, each vertex in $N_{\mathrm{cri}}$ is connected with half of the vertices of $N_{\mathrm{cri}}$.

In Figure \ref{fig:contact-graphical} we provide a graphical representation of $N_{\mathrm{con}}$. The size of each vertex is proportional to its degree. We used different colors to pinpoint the role of mobsters in the Mafia syndicate: in yellow we report the leaders of the organization (the so-called ``boss''). Green vertices represent {\em lieutenants}, i.e., the heads of a branch of a Mafia syndicate who commands a crew of soldiers (known as {\em picciotti}) and reports directly to the boss. Blue vertices represent actual mobsters, i.e., individuals who are known to be members of the syndacate. 
Blue vertices in the phone traffic network are not key network actors as they are spread all over the structure of the network, oftentimes in peripheral positions. In fact, both their position and ranking are often not prominent.

Figure \ref{fig:crimenet-graphical} shows the criminal network of dataset $N_{\mathrm{cri}}$. It is composed of 2590 links referring to relationships other than telephone-based contacts among the vertices of the network (for example but not only, complicity in a crime, acquaintance, police inspections, bank transactions, etc.) found by the prosecutors during the investigations.
$N_{\mathrm{cri}}$ includes the subset $I_{\mathrm{cri}}=\{ 13, 26, 84, 15, 76, 54 \}$ whose members are not present in $N_{\mathrm{con}}$. They are mobsters that were never tapped during investigations.
The structure is characterized by two clusters (clans) tied together by the subset $L_{\mathrm{cri}}=\left\lbrace 14, 15, 48, 49 \right\rbrace \subset N_{\mathrm{cri}}$ whose members are the so-called lieutenants (in green). As expected, the bosses (yellow vertices) of subset $B_{\mathrm{cri}}=\left\lbrace 100, 101, 102, 103, 104 \right\rbrace$ are situated between the two clusters, have a small number of links and at first look are not marginal.

Figure \ref{fig:aggregate-graphical} represents the aggregated network $A_{\mathrm{aggr}}$ which comprises the overall structure of the two networks in study. We highlighted the vertices representing the bosses $B_{\mathrm{cri}}$ (in yellow) together with the members of the subset $I_{\mathrm{cri}} \subset N_{\mathrm{cri}}$ not belonging to $N_{\mathrm{con}}$. The density of connections among the elements of $N_{\mathrm{cri}}$ changes the structure of network $N_{\mathrm{aggr}}$. Indeed, the two clusters of $N_{\mathrm{cri}}$ appears in the center, so that a core is formed.

From Table \ref{tab:CNmeasures} ---which summarizes information about the datasets--- and from the graphical representations shown in Figure \ref{fig:criminal-contact-network-graphical}, we can conclude that almost every member of criminal network $N_{\mathrm{cri}}$ under arrest is also a member of network $N_{\mathrm{con}}$.
Interestingly, telephone-based relationships among associates are rare in $N_{\mathrm{con}}$. Indeed only seven links were found, namely $C_{\mathrm{cri}}=\{(90,2),(104,87),(50,87),(28,87),(33,87),(19,87), (23,93)\}$.

\begin{figure*}[t!]
	\subfigure[]{
		\label{fig:criminal-contact}
		\begin{minipage}[tb]{0.32\textwidth}
			\centering
			\includegraphics[width=\textwidth]{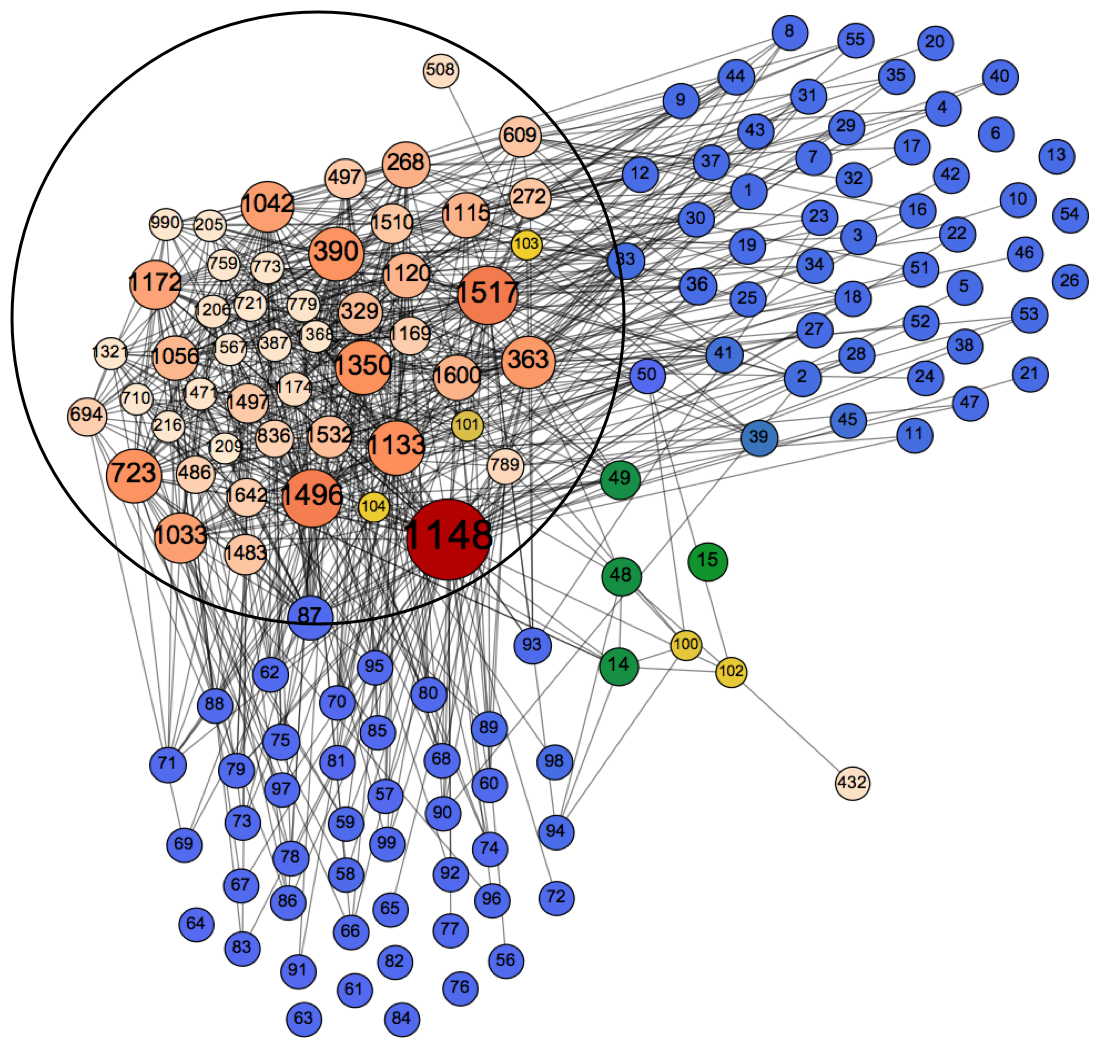}
		\end{minipage}}
	\subfigure[]{
		\label{fig:criminal-contact-egonet}
		\begin{minipage}[tb]{0.32\textwidth}
			\centering
			\includegraphics[width=\textwidth]{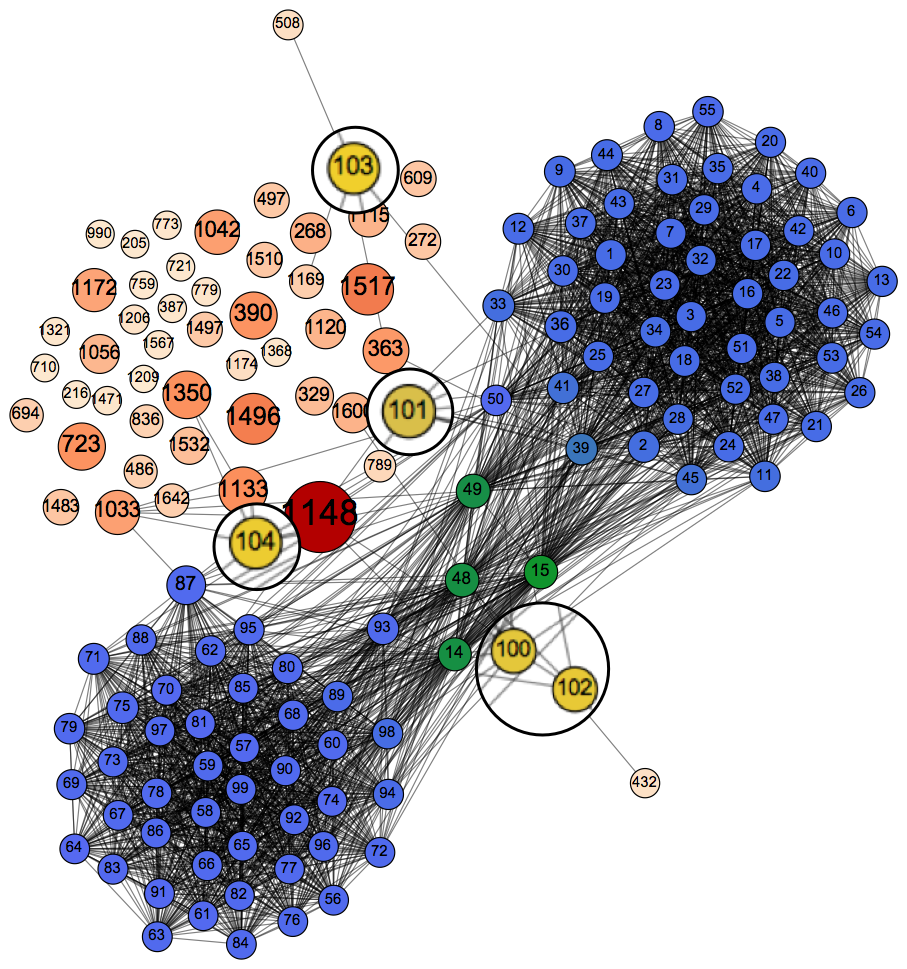}
		\end{minipage}}
	\subfigure[]{
			\label{fig:criminal-contacts-alls}
		\begin{minipage}[tb]{0.32\textwidth}
			\centering
			\includegraphics[width=\textwidth]{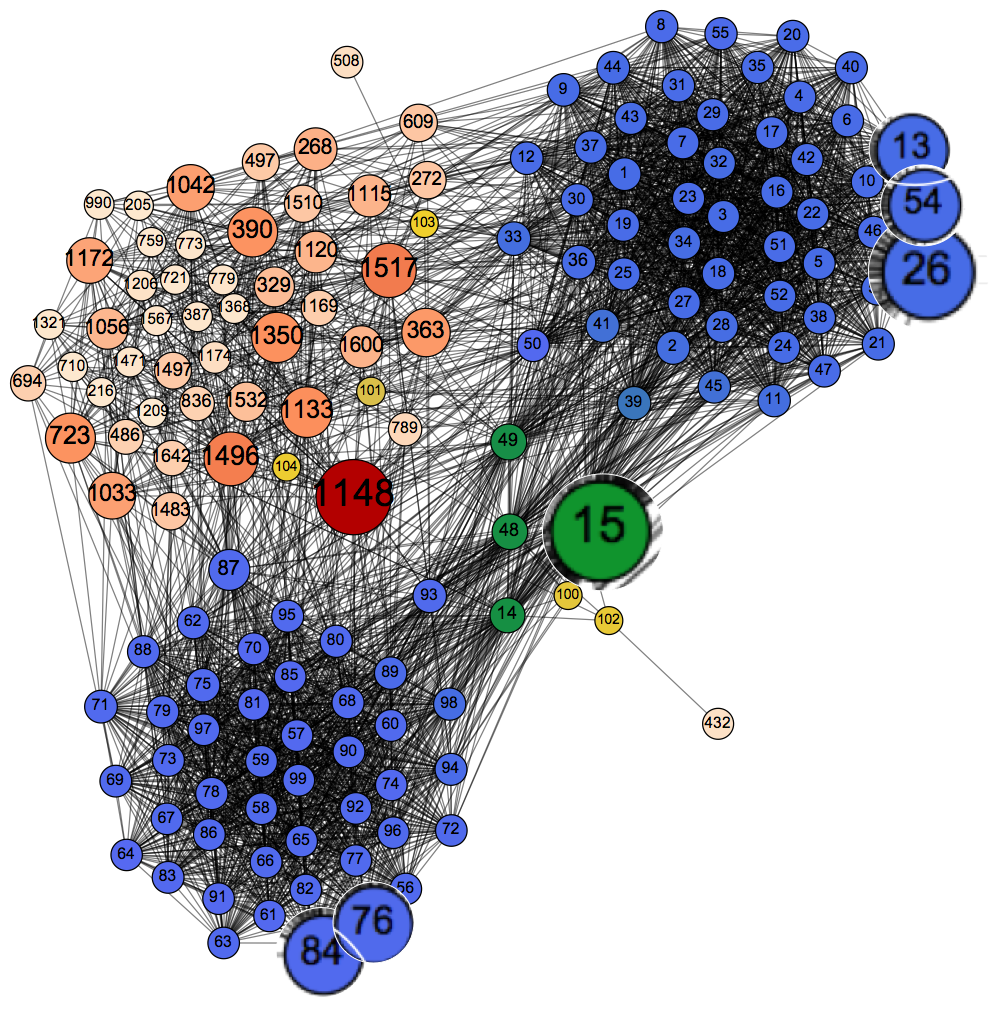}
		\end{minipage}}
	\subfigure[]{
		\label{fig:criminal-boss}
		\begin{minipage}[tb]{1.0\textwidth}
			\centering
			\includegraphics[width=\textwidth]{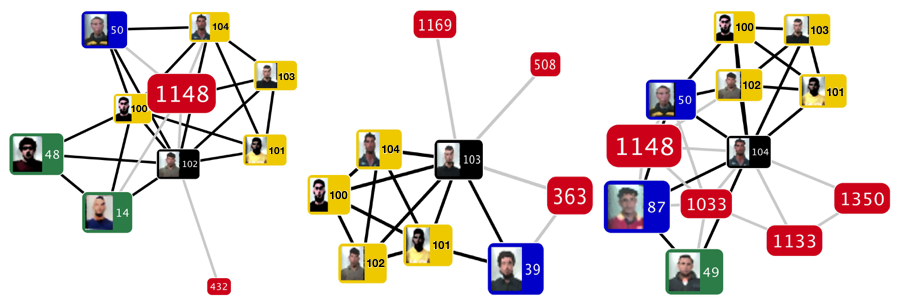}
		\end{minipage}}
\vspace{-0.3cm}
\caption{{\bf Panel (a):} We show all connections among the vertices belonging to $X_{cri}$ together with a subset of $N_{\mathrm{con}}$ having a high value of degree. {\bf Panel (b): } We show the edges of the network $N_{\mathrm{cri}}$ together with the edges connecting the elements of $B_{\mathrm{cri}}$ and $N_{\mathrm{con}}$. {\bf Panel (c): } We show all criminal and telephone-based connections of  network $N_{\mathrm{cri}}$ and we highlight (zoom) vertices of subset $I_{\mathrm{cri}}$. {\bf Panel (d): } We shown the egonets of bosses $\{102, 103, 104 \}$ filtered via the tool {\it LogAnalysis} \cite{catanese2013forensic}. The black lines represent edges of  set $E_{crim}$, the grey lines represent edges of  set $E_{con}$. Color codes: yellow vertices represent bosses, green vertices represent lieutenants, blue vertices represent associates. Red vertices denote members of the telephone-based network $N_{\mathrm{con}}$.}
\end{figure*}
It is known that associates are aware of investigative techniques and are inclined to minimize direct telephone-based communications. Direct communications are accomplished by intermediaries without a criminal record, above suspicion and unknown to law enforcement agencies.
This feature is clearly illustrated in Figure \ref{fig:criminal-contact} in which all telephone-based connections are shown among vertices $X_{cri}$ and a subset of most central vertices belonging to $N_{\mathrm{con}}$. In our opinion this is one of the most strategic elements to assure the resilience of a network. In this way a criminal network is not exposed to destabilizing attacks of law enforcement agencies because it manages a very limited number of telephone-based contacts among the members of the network. Nevertheless communications are still spread via elements that are not directly ascribable to the network.

Figure \ref{fig:criminal-contact-egonet} shows the connections of network $N_{\mathrm{cri}}$ and those among the members of $B_{\mathrm{cri}}$ and $N_{\mathrm{con}}$. In this case some of the most central vertices of the network $N_{\mathrm{con}}$ are implicated. The egonets of bosses $\{102, 103, 104\}$ are shown in Figure \ref{fig:criminal-boss}. Finally, Figure \ref{fig:criminal-contacts-alls}  shows the subgraph $N_{\mathrm{aggr}}$ which comprises of all the criminal and telephone-based connections of $N_{\mathrm{cri}}$ where we highlighted the vertices of the subset $I_{\mathrm{cri}}$.

\begin{figure*}[!t]
	\subfigure[]{
		\label{fig:egonets-inside}
		\begin{minipage}[tb]{0.52\textwidth}
			\centering
			\includegraphics[width=\textwidth]{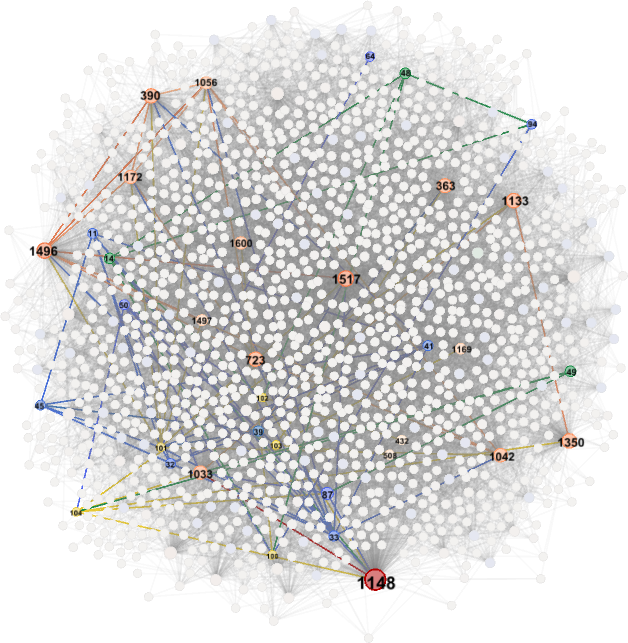}
		\end{minipage}}
	\subfigure[]{
		\label{fig:egonets-union}
		\begin{minipage}[tb]{0.48\textwidth}
			\centering
			\includegraphics[width=\textwidth]{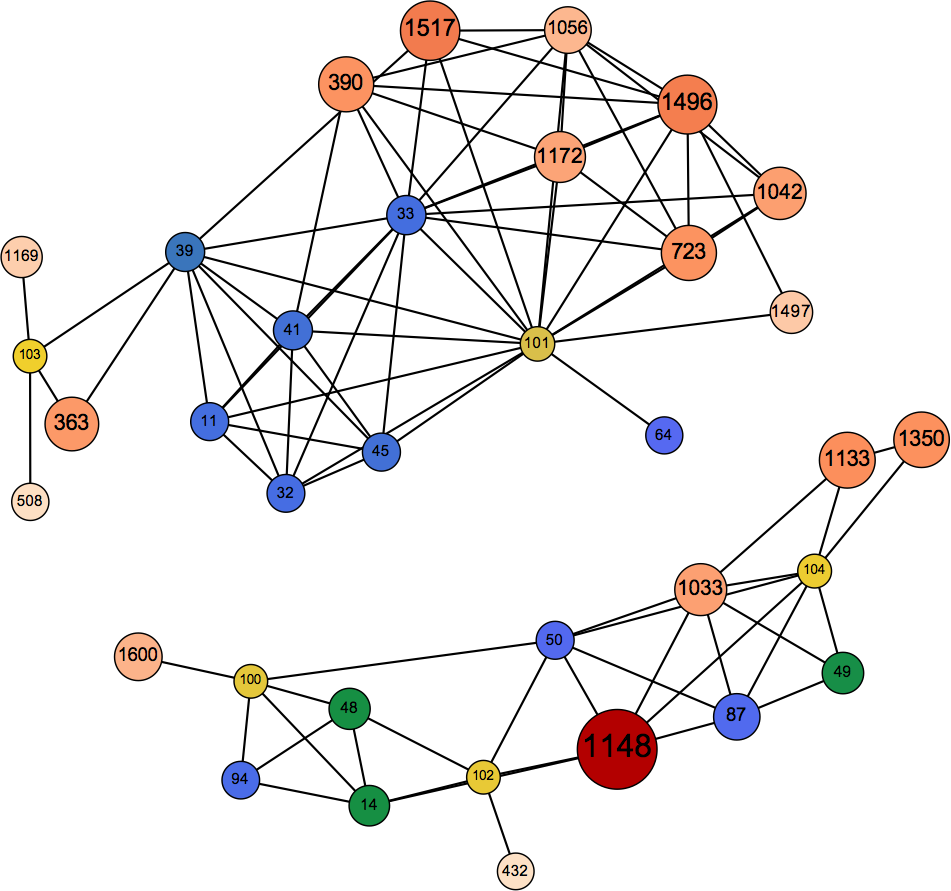}
		\end{minipage}}
\vspace{-0.3cm}
\caption{{\bf Left panel}: Network $A_{\mathrm{aggr}}$ in which are highlighted the egonets of the bosses $B_{\mathrm{cri}}$ of the criminal network. {\bf Right panel}: Subgraphs $B_{\mathrm{ego1}} \subset A_{\mathrm{aggr}}$ and $B_{\mathrm{ego2}} \subset A_{\mathrm{aggr}}$ of the egonets of the bosses obtained as the union of the egonets of every vertex of $B_{\mathrm{cri}}$.}
\label{fig:egonets}
\end{figure*}

The analysis of vertices of bosses and their position within the structure of the network gives an important insight in the study of resilience of a criminal network.
As we can see in Figure \ref{fig:egonets-inside}, bosses of the organization do not occupy important positions in the network $A_{\mathrm{aggr}}$. Nevertheless they are connected to the most important vertices in terms of degree.
Even in this case, relations among the most authoritative members of the organization are limited in time and amount. They manage the overall network indirectly via trusted people that not necessarily belong to the criminal network. In Figure \ref{fig:egonets-union} this concept is even more evident. Two subgraphs are shown which are obtained from the union of the egonets of the bosses of the set $B_{\mathrm{cri}}$, precisely of the subgraph $B_{\mathrm{ego1}}=\{101_{ego}\}\ \bigcup  \{103_{ego}\}$  e $B_{\mathrm{ego2}}=\{100_{ego}\} \bigcup \{102_{ego}\} \bigcup \{104_{ego}\}$ in which the bosses are connected to very few strategic vertices  to guarantee communications and flow of orders towards all the members of the criminal network. 
Direct connections among the bosses of the set $B_{\mathrm{ego1}}$ and the set $B_{\mathrm{ego2}}$ within the network $A_{\mathrm{aggr}}$ (the bosses of the two groups never had telephone-based communications among them, had meetings escaping the investigations, were never charged of the same crime, never left evidence of bank transactions, etc.). This is a further element of resilience of the criminal network: the removal of a vertex from a subgroup has no consequences on the other subgroup. Nevertheless the bosses are tightly tied and occupy the uppermost position in the criminal organization. This is why in Figure \ref{fig:criminal-boss} we decided to include even missing relations (not present in the datasets), in order to increase the meaning of the visualization.

\subsection{Analysis of the structural properties of contact and criminal networks}
The next step of our analysis consists of studying the structural properties of $N_{\mathrm{cri}}$ and $N_{\mathrm{con}}$.
To perform our analysis we considered two main parameters:

{\bf Degree Distribution}. Given a vertex $i$ in $N_{\mathrm{con}}$ (resp., $N_{\mathrm{cri}}$), we compute its degree $k_i$ in $N_{\mathrm{con}}$ (resp., $N_{\mathrm{cri}}$).
From the analysis of the degree distribution it is possible to check if there are vertices in $N_{\mathrm{con}}$ (resp., $N_{\mathrm{cri}}$) which are much more connected than other or, vice versa, if the number of connections of each individual is roughly the same.

{\bf Average Clustering Coefficient}. Given a vertex $i$ in $N_{\mathrm{con}}$ (resp., $N_{\mathrm{cri}}$), we define the {\em neighborhood of level 1} $\mathcal{N}(i)$ associated with $i$ as the set of vertices which are adjacent to $i$ in $N_{\mathrm{con}}$ (resp., $N_{\mathrm{cri}})$.\footnote{The vertex $i$ is not considered in $\mathcal{N}(i)$.}
The {\em average clustering coefficient} $\mathtt{ACC}_i$ of $i$ is defined as follows:

$$
ACC_i = \frac{2\times |\{\langle v, w \rangle: v \in \mathcal{N}(i), w \in \mathcal{N}(i)\}|}{k_i \times \left(k_i -1\right)}
$$
Here, $\{\langle v, w \rangle: v \in \mathcal{N}(i), w \in \mathcal{N}(i)\}$ is the set of pairs of vertices $v$ and $w$ which are connected each other and, simultaneously, are both connected to $i$.
The triplet formed by vertices $i$, $v$ and $w$ is also called {\em closed triplet} and, therefore, $ACC_i$ measures the number of closed triplets having a vertex in $i$ out of the total number of triplets of vertices that contain $i$.
The $ACC_i$ ranges in $[0,1]$ and if it is nearly equal to 1 then the neighbors of $i$ tend to form a large number of triangles which favors the spreading of information.

\begin{figure}
\centering
\includegraphics[width=0.6\textwidth]{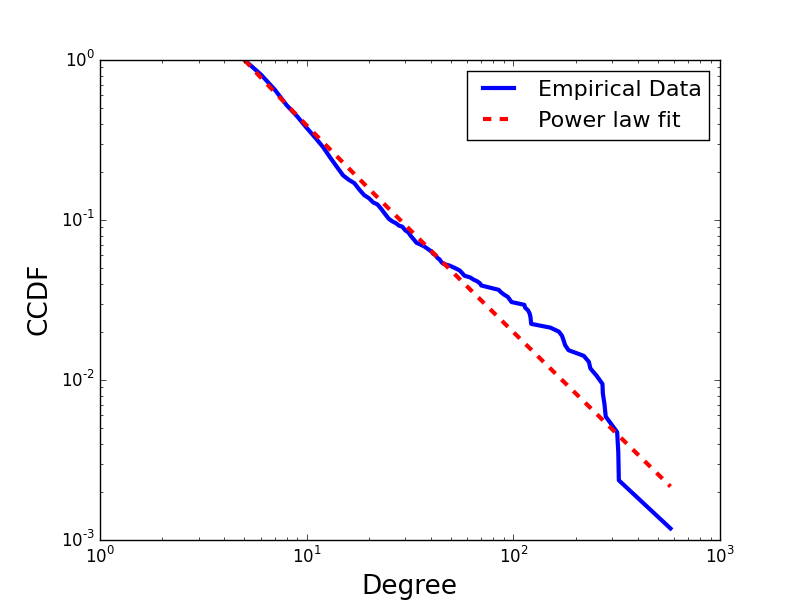}
\label{fig:contact-network-degree}
\caption{The CCDF associated with the degree distribution $k_i$ in $N_{\mathrm{con}}$. We used a $\log$-$\log$ scale and, in the same plot, we report the power law distribution best fitting the experimentally observed data.}
\end{figure}

We begin our study by discussing the vertex degree distribution in both $N_{\mathrm{con}}$ and $N_{\mathrm{cri}}$.
As for $N_{\mathrm{con}}$, we plotted the Cumulative Complementary Distribution Function CCDF which specifies, for a fixed threshold $\overline{k}$, the probability that a randomly selected vertex has degree greater than $\overline{k}$.
We displayed the CCDF in Figure \ref{fig:contact-network-degree} on a log-log scale.

Similarly to many other socio-technical systems, contacts among individuals in $N_{\mathrm{con}}$ are rather sparse and unevenly distributed, with few vertices capturing most of the edges in $N_{\mathrm{con}}$. We used the statistical tool described in \cite{alstott2014powerlaw} and we found that the degree distribution followed a power law with $\alpha = -2.5$ ($p$-value $< 10^{-5}$).
		
\begin{figure}
\centering
\includegraphics[width=0.6\textwidth]{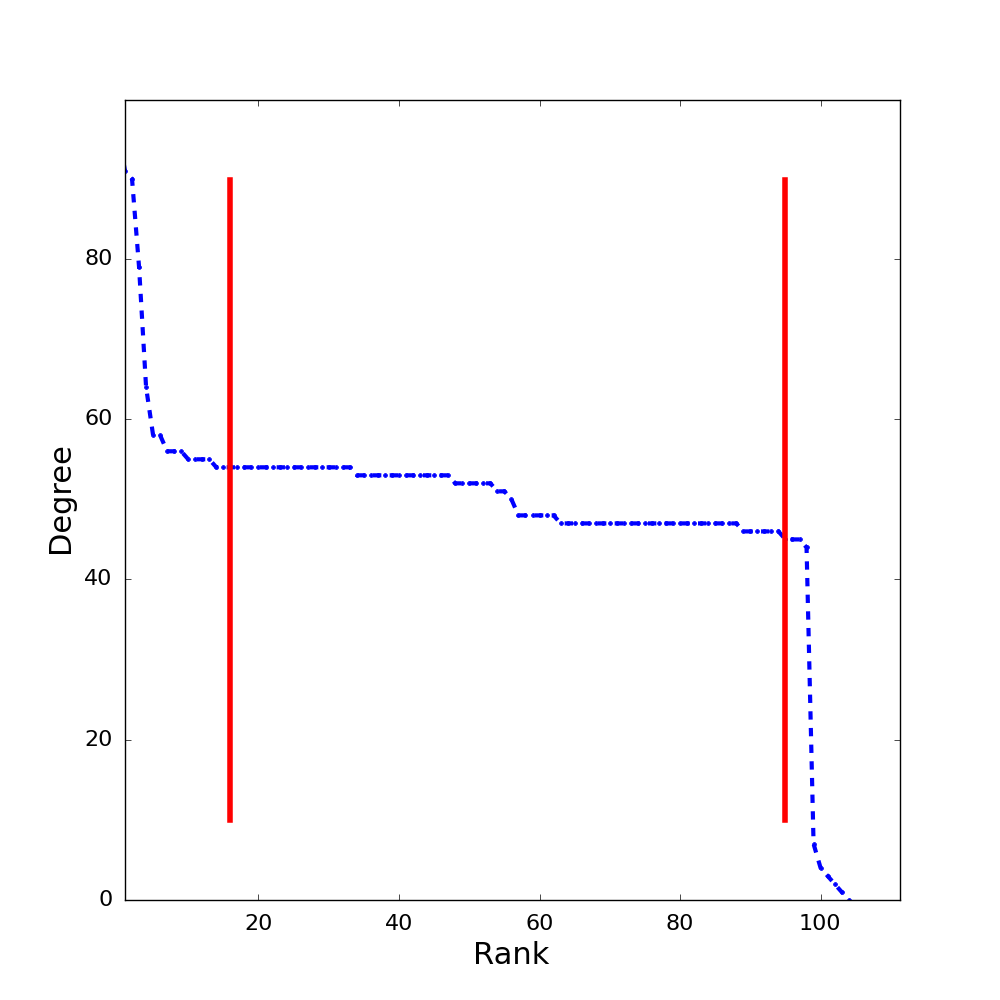}
\label{fig:criminal-network-degree}
\caption{We report the degree of each vertex vs. its rank: the vertex with rank $\ell$ is the vertex having the $\ell$-th largest degree. We split vertices on the basis of their degree and we obtained three classes, namely {\em Group A} ($0 \ k_i \leq 15$), {\em Group B} ($15 < k_i \leq 85$) and {\em Group C} ($k_i > 85$).}
\end{figure}

Because of $N_{\mathrm{con}}$ is quite dense, we adopted a different graphical procedure to investigate vertex degree distribution. We ranked vertices in $N_{\mathrm{cri}}$ on the basis of their degree: in this way the $\ell$-th ranked vertex is the vertex showing the $\ell$-th largest degree (see Figure \ref{fig:criminal-network-degree}).

We noticed few individuals who were connected to almost all other individuals in $N_{\mathrm{con}}$ but there was also a small number of individuals who were connected to few individuals and just one of them was isolated (i.e., there was a vertex with degree 0).
The vast majority of vertices in $N_{\mathrm{con}}$ had a degree ranging from 15 to 85.

In $N_{\mathrm{con}}$ we found few individuals who are well connected with many other individuals but the largest part of vertices shows a low degree. In contrast, in $N_{\mathrm{cri}}$ there were only 15 individuals with degree less than 40 and only 17 individuals with degree larger than 55. This suggests the possibility to partition individuals in $N_{\mathrm{con}}$ in three disjoint classes on the basis of their degree, namely: {\em (i) Class A}, if $k_i \leq 15$, {\em (ii) Class B}, if $15 < k_i \leq 85$ and {\em (i) Class C}, if $k_i > 85$.
With the aid of police officers, we observed that individuals belonging to {\em Class A} did not had leadership roles in the gang but they often acted as intermediaries. Surprisingly enough, all the heads of the gang were members of {\em Class C} and this depends on the fact that leaders in $N_{\mathrm{cri}}$ are aware of risks and, therefore, they are in touch with just an handful of gang associates.
The density of $N_{\mathrm{cri}}$ is likely to depend on the nature of many crimes: activities like drug trafficking or gambling impose mobsters to organize into gangs and coordinate their actions. This implies that the resulting network must be dense and it well explains why the average degree is roughly 50, i.e., any mobster is connected with half of the members of $N_{\mathrm{cri}}$.

\begin{figure}[!t]
	\centering
		\includegraphics[width=1.0\columnwidth]{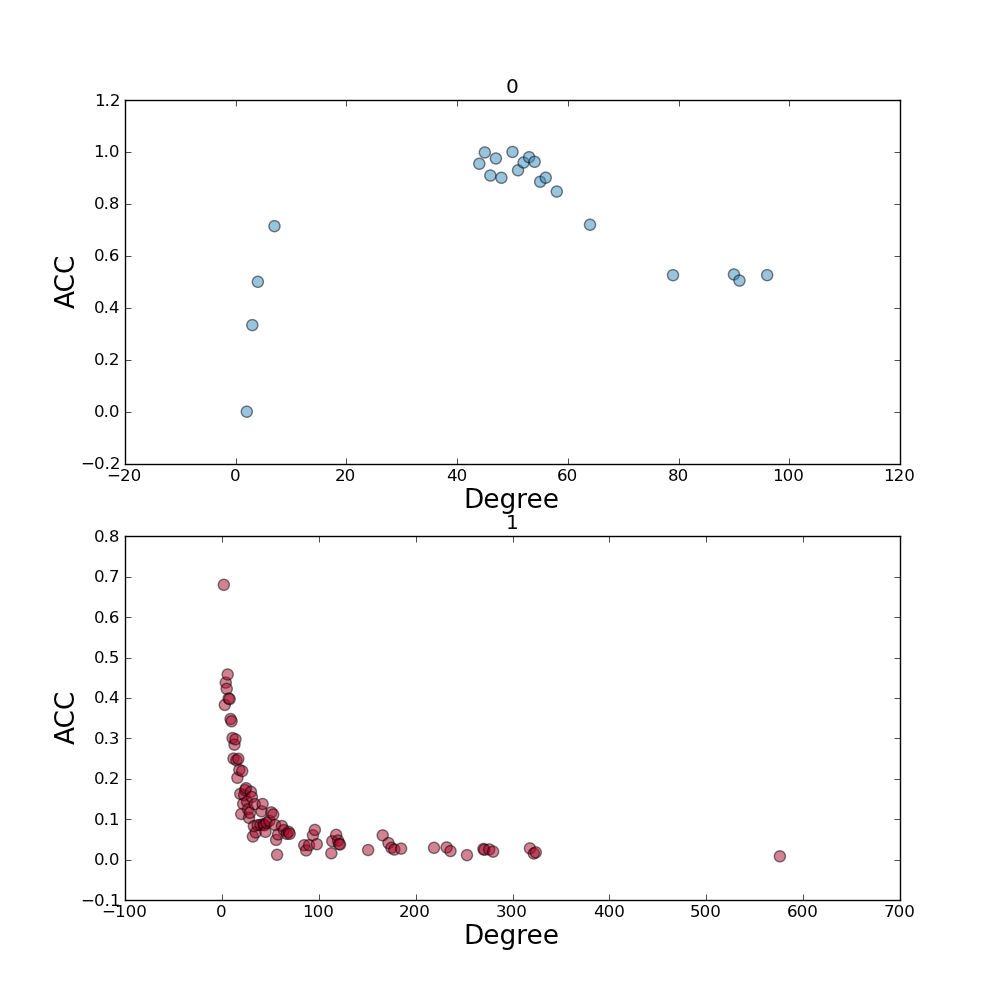}
		\label{fig:acc-vs-degree}
\caption{{\bf Top}: Average Clustering Coefficient as function of $k_i$ in $N_{\mathrm{con}}$. {\bf Bottom}: Average Clustering Coefficient as function of $k_i$ in  $N_{\mathrm{cri}}$. }
\end{figure}

We continue our analysis by focusing on the structure of the social relationships of a given individual in both $N_{\mathrm{con}}$ an $N_{\mathrm{cri}}$.
Figure \ref{fig:acc-vs-degree} shows the values of Average Clustering Coefficient as function of vertex degree for both Contact and Criminal Networks.
In case of $N_{\mathrm{cri}}$, $\mathtt{ACC}$ features generally low values and it is monotonically decreasing with the degree $k_i$.
We notice that $\mathtt{ACC}$ in $N_{\mathrm{cri}}$ is always bigger than 0.6 which is a surprisingly large value: many socio-technical systems and Web platforms like Facebook or MSN messenger, in fact, generally feature a value of $\mathtt{ACC}$ in the range of $0.01-0.14$ \cite{ugander2011anatomy, catanese2011crawling}. In addition, $\mathtt{ACC}$ achieves its peak for {\em Class B} users. Such a result can be paired up with our previous discussion: in $N_{\mathrm{cri}}$ there is a large fraction (which account for roughly 90\% of the whole population) of individuals with degree ranging from 40 to 60 and, at the same time, the contacts of these users are themselves well-connected each other.
		
The abundance of triangles in $N_{\mathrm{cri}}$ depends on the normative structure governing syndicates. In fact, past testimony \cite{maas1968valachi} as well as documents found during the arrest of  Mafia boss Salvatore Lo Piccolo showed that one of the first rule in the Mafia decalogue was as follows: ``No one can present himself directly to another of our friends. There must be a third person to do it.''\footnote{See \url{http://news.bbc.co.uk/2/hi/europe/7086716.stm}}

\section{Resilience of criminal and contact networks in a parallel police operation}
\label{sec:resilience-parallel}
		
In this section we aim at studying the resilience of the contact and criminal networks at our disposal to both random and targeted attacks.
The resilience of a contact/criminal network is a crucial parameter to quantify the ability of a Mafia syndicate to react to the arrest of some of its members.
More in general, Mafia syndicates tend to structure themselves in an way that is resilient to police operations aiming at hindering or inhibiting temporarily or permanently the functions of any specific member of the organization.

To assess network robustness in presence of the dismissal of some of its members we rely on previous studies \cite{albert2000error, broder2000graph} discussed in Section \ref{sub:network-robusteness}, and  we consider two parameters: {\em (i)} the size of the largest strongly connected component $\mathtt{SCC}$ and {\em (ii)} the average path length $\mathtt{APL}$.

A large value of $\mathtt{SCC}$ implies that a pair of arbitrary selected individuals in  $N_{\mathrm{con}}$ (resp., $N_{\mathrm{cri}}$) is able to find a path (going through other individuals) along which a message can be routed. Relatively small values of $\mathtt{APL}$ imply that, on average, a user has to go through a short chain of intermediaries to get in touch with any other individual and, in the crime context, a quick flow of information is a crucial parameter to establish the survival of the organization itself.

We considered two attack strategies:
		
\begin{itemize}		
\item {\em Random attack strategy}. We selected, uniformly at random, a fraction $f$ of vertices from $N_{\mathrm{con}}$ ($N_{\mathrm{cri}}$) and removed them along with their incident connections. We then measured the corresponding variation of $\mathtt{SCC}$ and $\mathtt{APL}$.
To produce statistically robust results, we ran the procedure described above 100 times and computed the average of $\mathtt{SCC}$ and $\mathtt{APL}$.
In our experiment $f$ varied from $1\%$ to $25\%$.
			
\item {\em Targeted Attacks with DC, BC and CC strategies}. We computed the centrality of each vertex by applying one of the three centrality indices introduced in Section \ref{sub:centrality-scores}, i.e., Degree Centrality ($DC$), Betwenness Centrality ($BC$) and Closeness centrality ($CC$). For each centrality index, we sorted vertices on the basis of their centrality score and deleted a fraction $f$ of them from $N_{\mathrm{con}}$ ($N_{\mathrm{cri}}$) along with their connections. We then measured the corresponding variation of $\mathtt{SCC}$ and $\mathtt{APL}$. Once again, $f$ varied from $1\%$ to $25\%$.
\end{itemize}
		
\begin{figure*}[t!]
	\subfigure[]{
	\label{fig:contact-wcc}
		\begin{minipage}[tb]{0.49\textwidth}
			\centering
			\includegraphics[width=\textwidth]{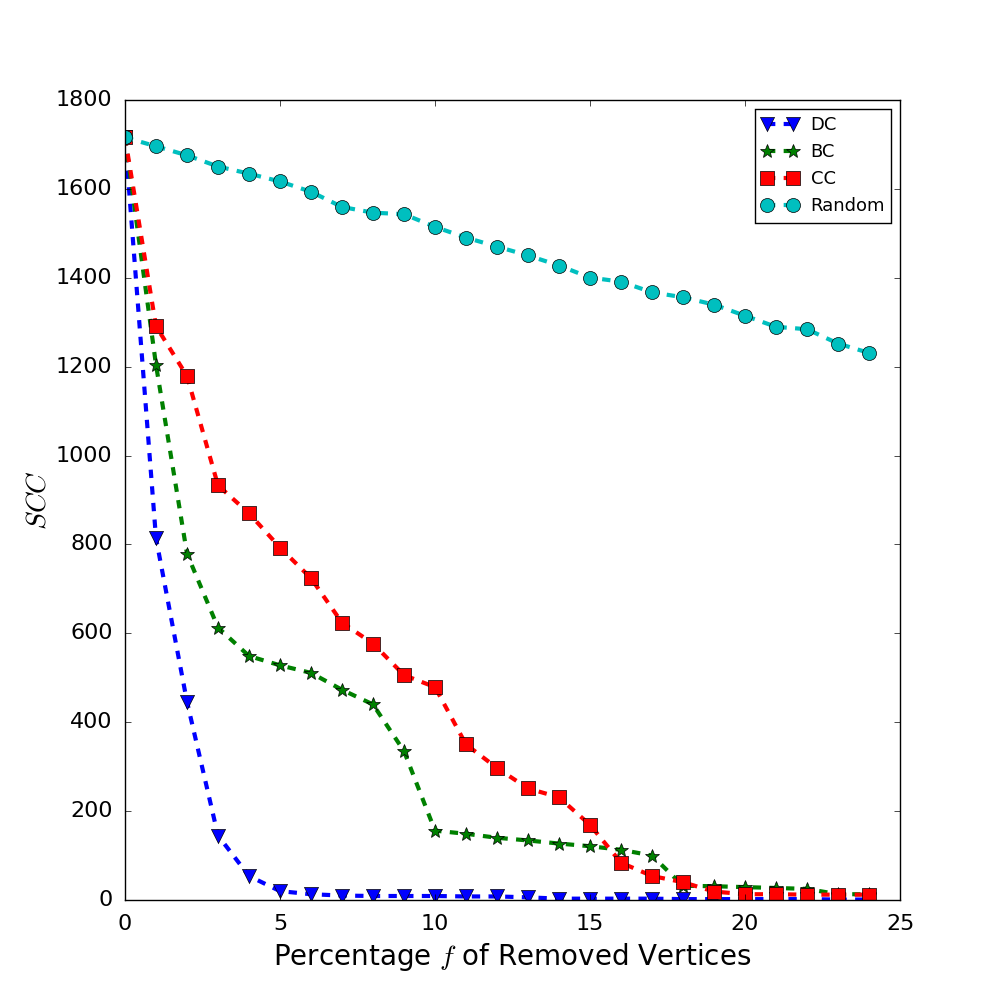}
		\end{minipage}}
	\subfigure[]{
	\label{fig:criminal-wcc}
		\begin{minipage}[tb]{0.49\textwidth}
			\centering
			\includegraphics[width=\textwidth]{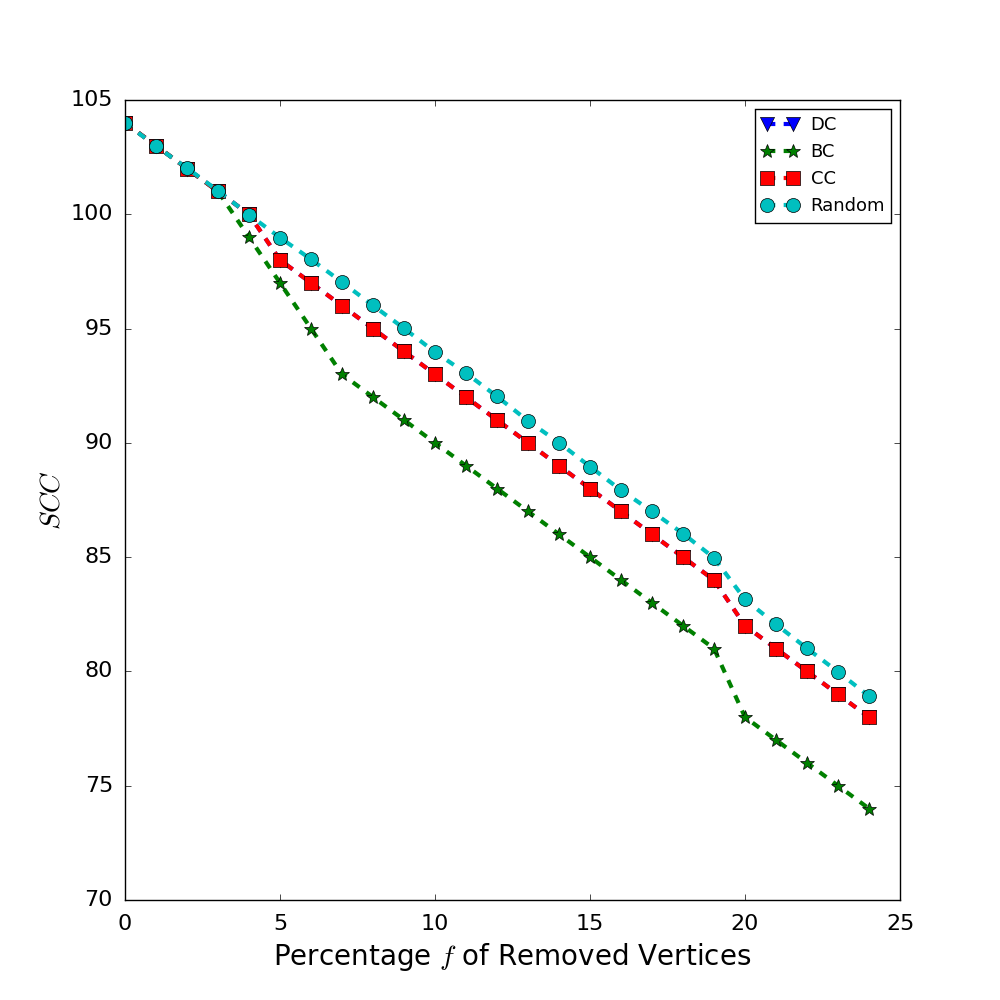}
		\end{minipage}}
\vspace{-0.3cm}
\caption{{\bf Left panel}: $\mathtt{SCC}$ vs. the fraction $f$ of removed vertices in $N_{\mathrm{con}}$ in case of parallel police operation. {\bf Right panel}: $\mathtt{SCC}$ vs. the fraction $f$ of removed vertices in $N_{\mathrm{cri}}$ in case of parallel police operation.}
\label{fig:wcc-disruption}
\end{figure*}

The outcome of our experiments are graphically reported in Figures \ref{fig:contact-wcc} -\ref{fig:criminal-wcc}.
		
As for $\mathtt{SCC}$, we observe that targeted attacks are able to quickly destroy the strongly connected component in case of $N_{\mathrm{con}}$. In particular, from Figure \ref{fig:contact-wcc}, $DC$ has the most disruptive effect on $\mathtt{SCC}$ and the removal of less than $5\%$ of the most central vertices is enough to completely destroy the largest connected component.
Random attacks yield a linear decrease in $\mathtt{SCC}$ and if $f$ shifts from $5\%$ to $25\%$, then $\mathtt{SCC}$ decreases of about $24\%$.
$BC$ and $CC$ are respectively the second and third most effective strategies, however they require, on average, a removal of between 15\% and 20\% if vertices to effectively disrupt $\mathtt{SCC}$.

Different conclusions can be drawn if we focus on $N_{\mathrm{cri}}$ (see Figure \ref{fig:criminal-wcc}). Such a network exhibits, in fact, an exceptional degree of robustness and, independently of the centrality index we decided to adopt, we observe that $\mathtt{SCC}$ always decreases in a linear fashion.
Here, $BC$ yields the largest decrease in $\mathtt{SCC}$ and the lines associated with $DC$ and $CC$ mostly overlap.
This result illustrates that there are no obvious targeted strategies that effectively disrupt a criminal network, at least by using parallel police operations.
		
\begin{figure*}[t!]
	\subfigure[]{
		\label{fig:contact-apl}
			\begin{minipage}[tb]{0.48\textwidth}
			\centering
			\includegraphics[width=\textwidth]{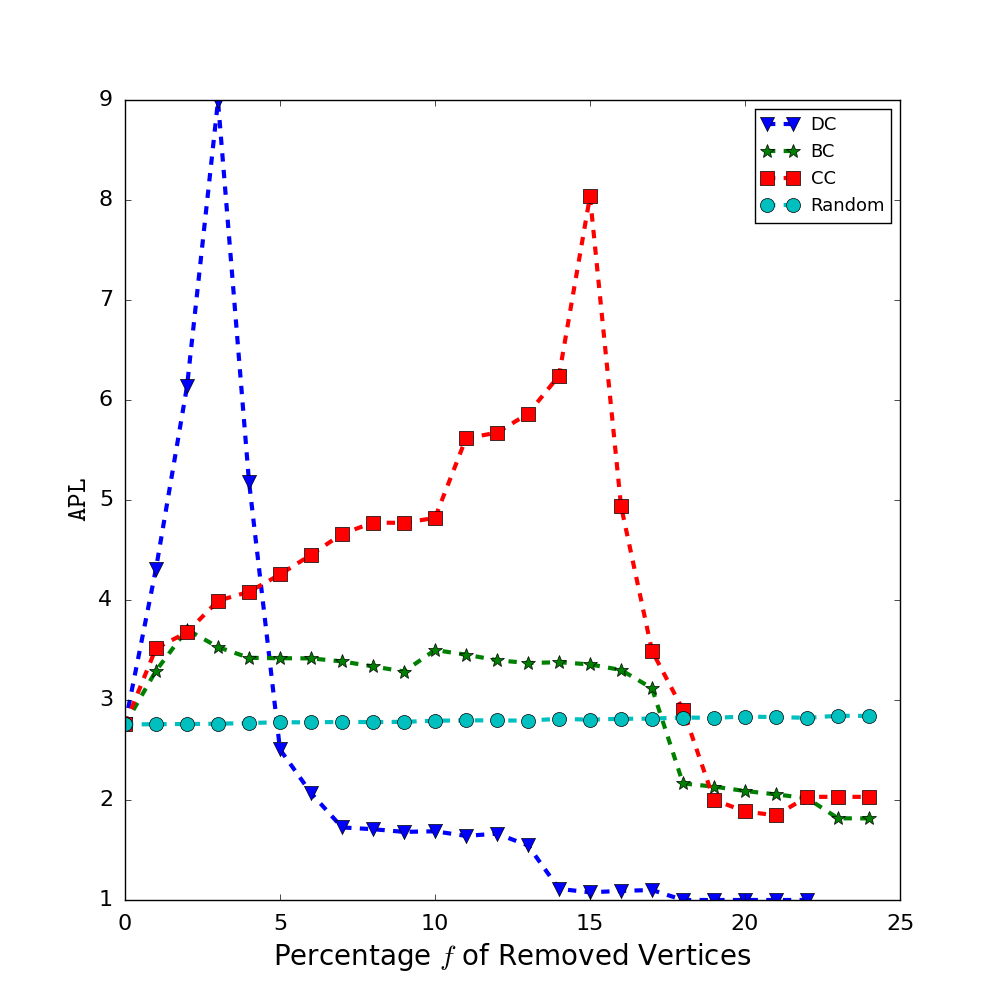}
			\end{minipage}}
	\subfigure[]{
		\label{fig:criminal-apl}
			\begin{minipage}[tb]{0.48\textwidth}
			\centering
			\includegraphics[width=\textwidth]{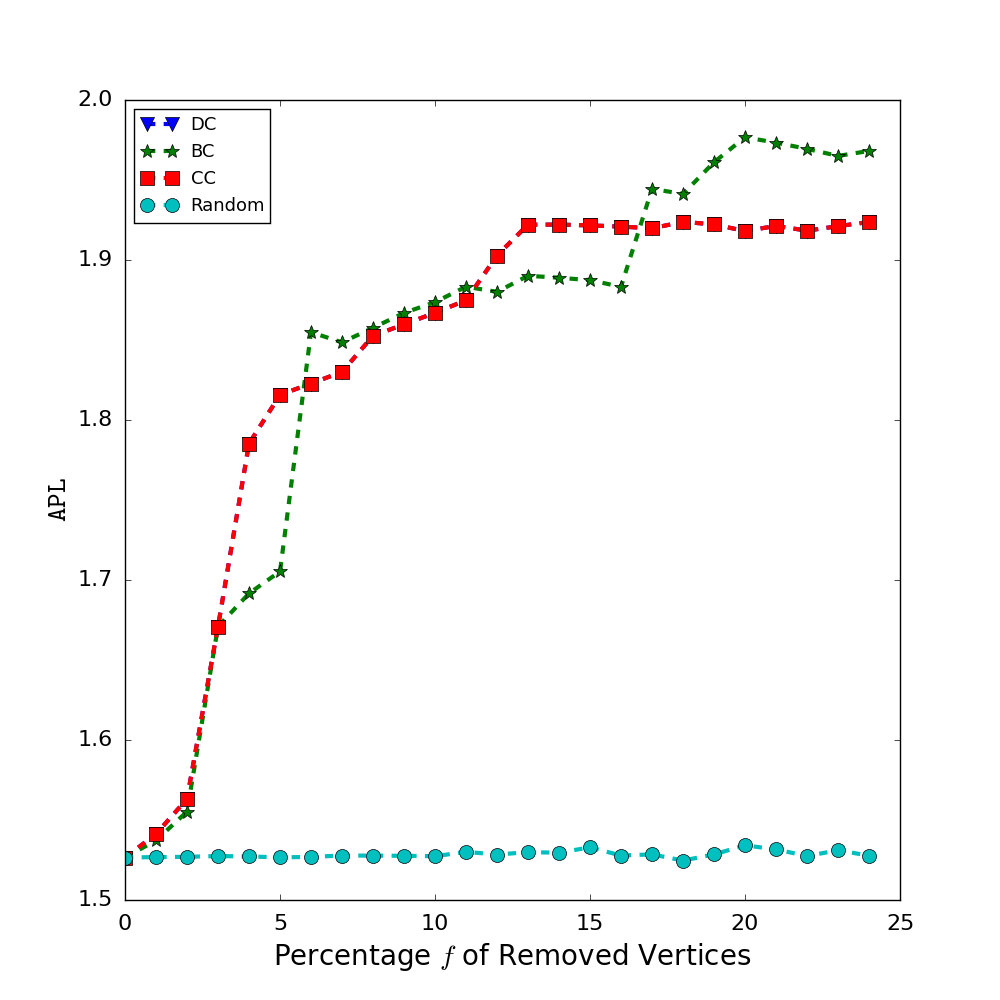}
			\end{minipage}}
\vspace{-0.3cm}
\caption{{\bf Left panel}: $\mathtt{APL}$ vs. the fraction $f$ of removed vertices in $N_{\mathrm{con}}$ in case of parallel police operation. {\bf Right panel}: $\mathtt{APL}$ vs. the fraction $f$ of removed vertices in $N_{\mathrm{cri}}$ in case of parallel police operation.}
\label{fig:apl-disruption}
\end{figure*}

As a further experiment, we studied the variation of $\mathtt{APL}$ in $N_{\mathrm{con}}$ and $N_{\mathrm{cri}}$ when an increasing fraction
of vertices was deleted from these two graphs under both random and targeted attacks (see Figures \ref{fig:contact-apl} and \ref{fig:criminal-apl}).
								
We observe that random attacks are ineffective to increase the value of $\mathtt{APL}$ in case of $N_{\mathrm{con}}$.

Indeed, in $N_{\mathrm{con}}$ the most effective strategy is, once again, $DC$: we need to neutralize only the top 2\% vertices from $N_{\mathrm{con}}$ to nearly triplicate the $\mathtt{APL}$. Yet using $DC$, if $f > 4\%$  the contact network breaks into separate components.
Analogous observations hold if we use $BC$ and $CC$ to score vertices even if the breaking 
point roughly occurs again with $f = 15 - 20\%$.

Random attacks are, {\em de facto}, ineffective in augmenting $\mathtt{APL}$ in 
$N_{\mathrm{cri}}$.

Our experiments suggest that, in $N_{\mathrm{con}}$, it is sufficient to neutralize a small fraction of vertices (around $5-7\%$ if the $DC$ strategy is adopted)
to significantly reduce $\mathtt{SCC}$ and, simultaneously, increase $\mathtt{APL}$.
In contrast, due to its high density of crime ties, $N_{\mathrm{cri}}$ is much more resilient and therefore it is able to effectively react to targeted attacks.

\section{Resilience of criminal and contact networks in a sequential police operation}
\label{sec:resilience-sequential}
											
We conclude our study by focusing on a different type of police operation that we call {\em sequential scenario}. In a sequential police operation, we
wish to study how a criminal organization is able to re-organize itself when one (or more of its members) are neutralized.
We suppose that mobsters are arrested one-by-one and we measure how each arrest impacts  $\mathtt{SCC}$ and $\mathtt{APL}$. This leads us to design the following experimental procedure:  {\em (i)} We select a vertex according to the three centrality indices $DC$, $BC$ and $CC$ presented in Section \ref{sub:centrality-scores}; {\em (ii)} We neutralize the selected vertex by deleting it along with its incident connections. {\em (iii)} We calculate  $\mathtt{SCC}$ and $\mathtt{APL}$ on the network obtained at the end of Step {\em (ii)}. Steps {\em (i)}-{\em (iii)} are repeated until the top 30\% vertices of $N_{\mathrm{con}}$ (resp., $N_{\mathrm{cri}}$) are neutralized.
											
\begin{figure*}[t!]
	\subfigure[]{
		\label{fig:contact-wcc-sequential}
		\begin{minipage}[tb]{0.49\textwidth}
			\centering
			\includegraphics[width=\textwidth]{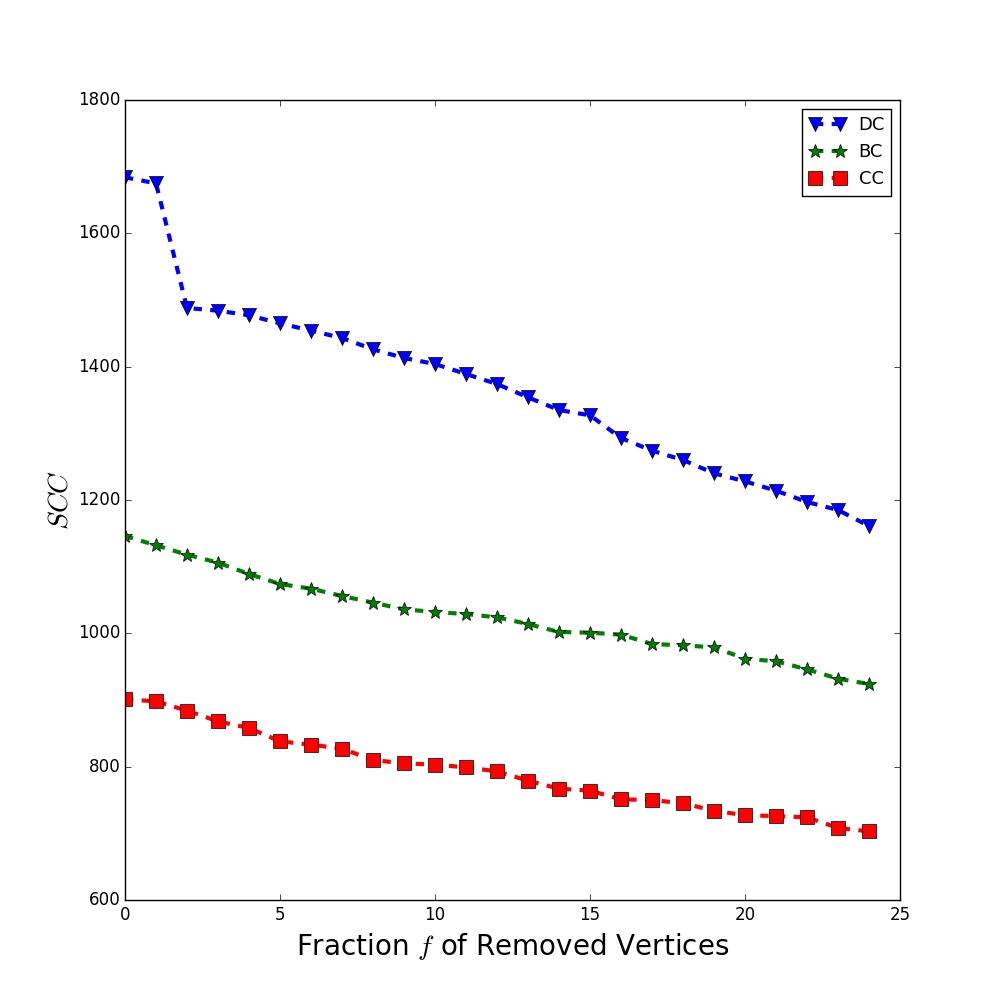}
		\end{minipage}}
	\subfigure[]{
		\label{fig:criminal-wcc-sequential}
		\begin{minipage}[tb]{0.49\textwidth}
			\centering
			\includegraphics[width=\textwidth]{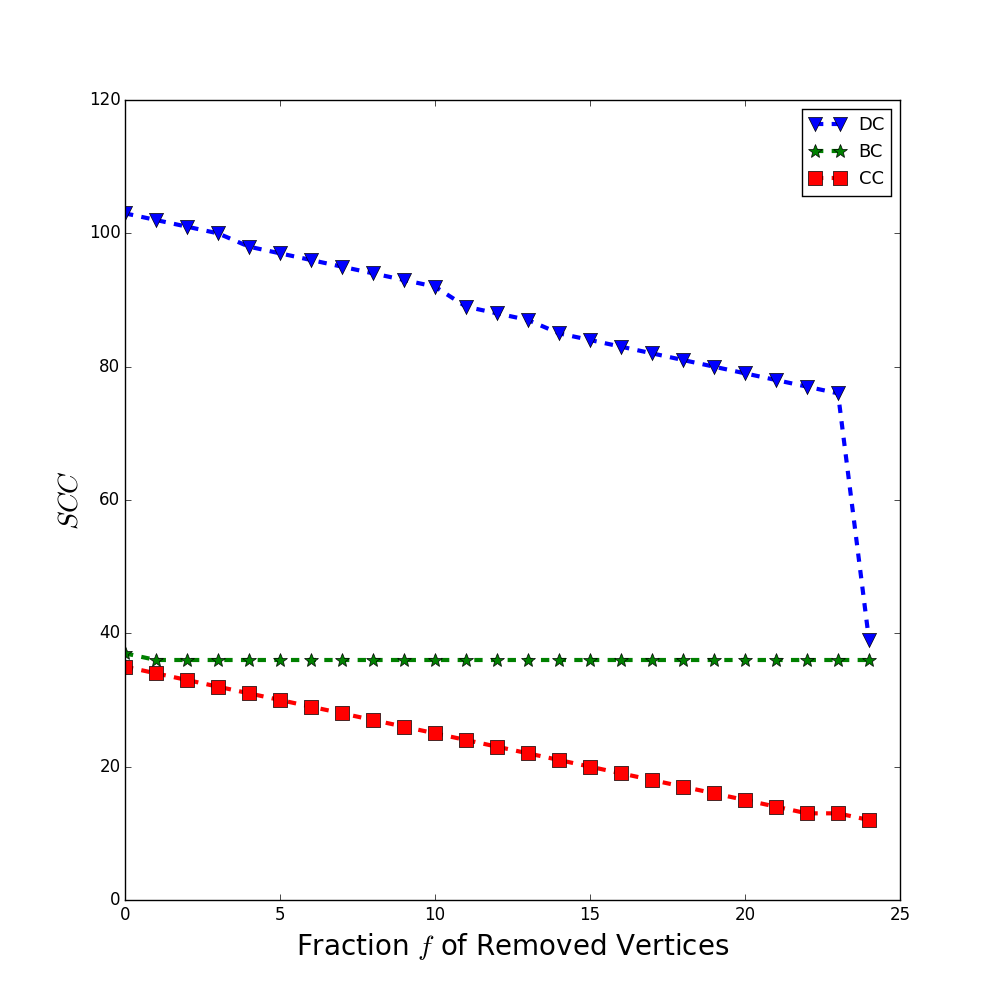}
		\end{minipage}}
\vspace{-0.3cm}
\caption{{\bf Left panel}: $\mathtt{SCC}$ vs. the fraction $f$ of removed vertices in $N_{\mathrm{con}}$ in case of a sequential police operation. {\bf Right panel}: $\mathtt{SCC}$ vs. the fraction $f$ of removed vertices in $N_{\mathrm{cri}}$ in case of a sequential police operation.}
\label{fig:wcc-disruption-sequential}				
\end{figure*}

\begin{figure*}[t!]
	\subfigure[]{
		\label{fig:contact-apl-sequential}
		\begin{minipage}[tb]{0.49\textwidth}
			\centering
			\includegraphics[width=\textwidth]{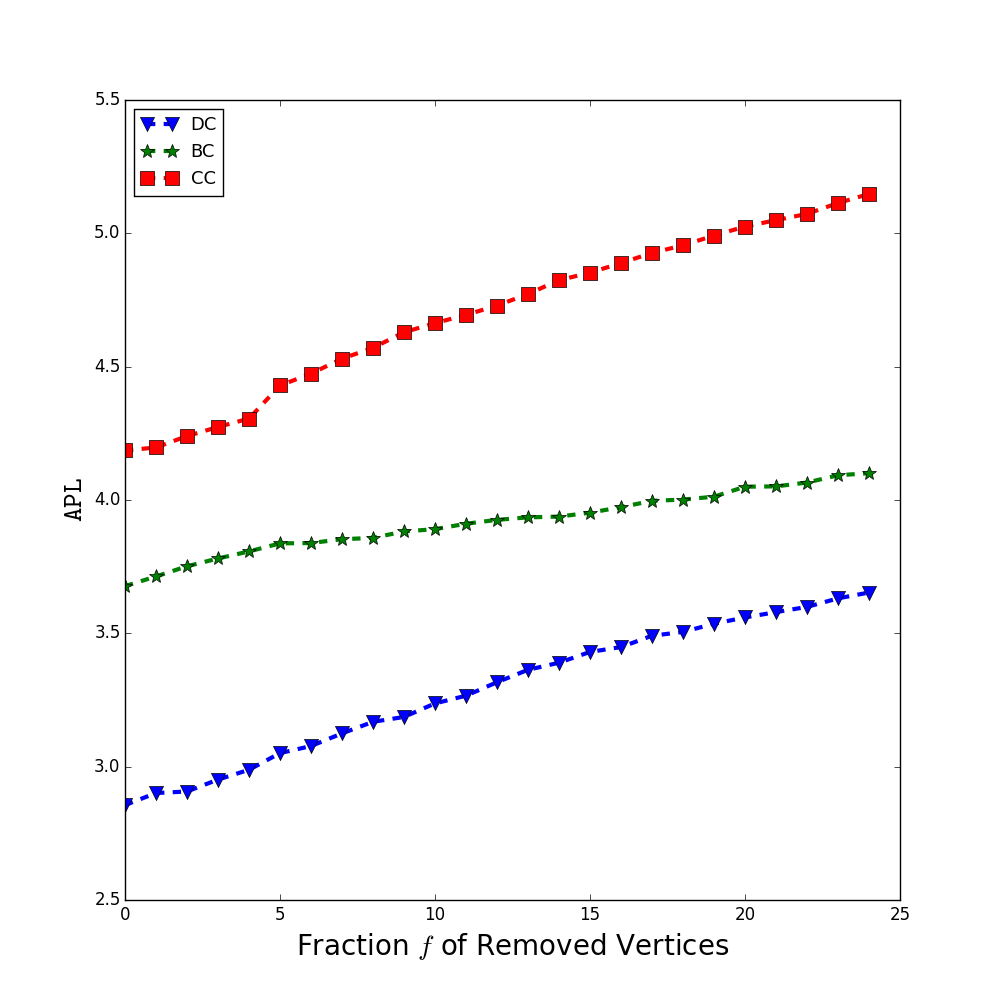}
		\end{minipage}}
	\subfigure[]{
		\label{fig:criminal-apl-sequential}
		\begin{minipage}[tb]{0.49\textwidth}
			\centering
			\includegraphics[width=\textwidth]{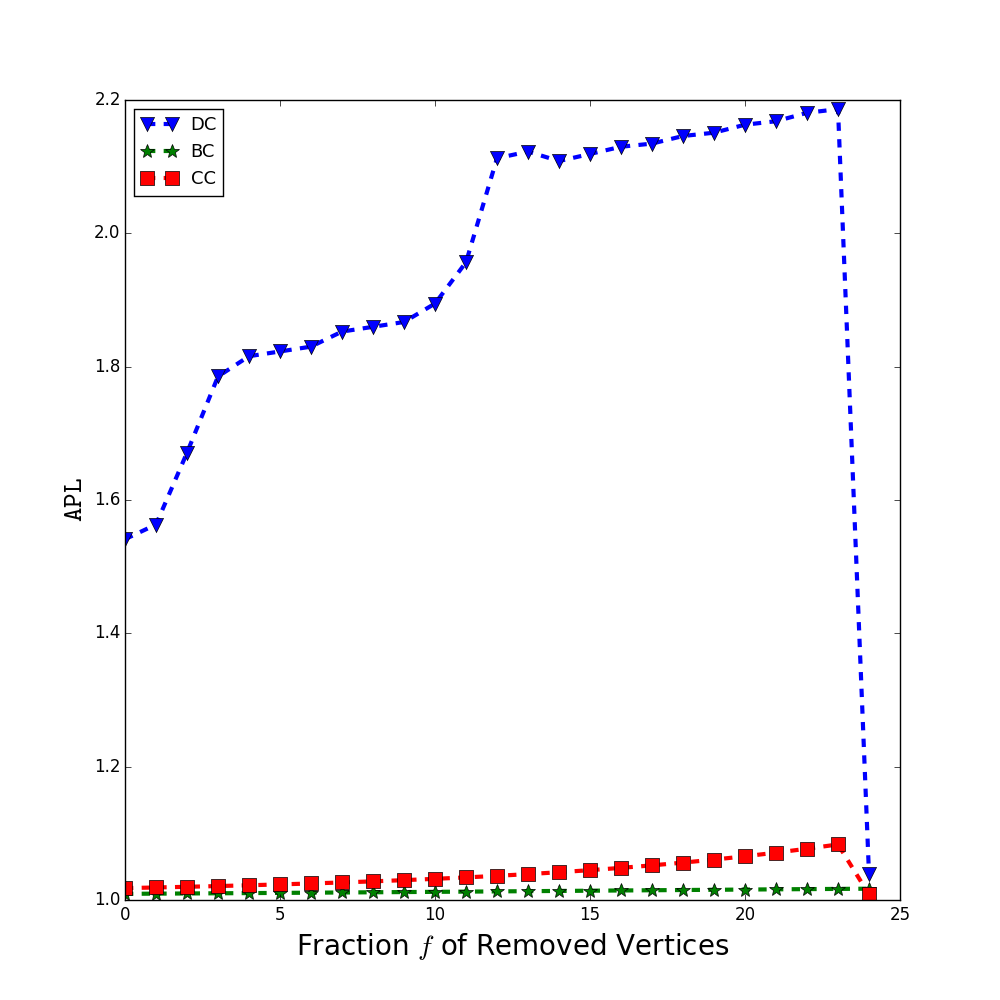}
		\end{minipage}}
\vspace{-0.3cm}
\caption{{\bf Left panel}: $\mathtt{APL}$ vs. the fraction $f$ of removed vertices in $N_{\mathrm{con}}$ in case of a sequential police operation. {\bf Right panel}: $\mathtt{APL}$ vs. the fraction $f$ of removed vertices $N_{\mathrm{cri}}$ in case of a sequential police operation.}
\label{fig:apl-disruption-sequential}
\end{figure*}

In Figures \ref{fig:contact-wcc-sequential}-\ref{fig:criminal-wcc-sequential} we plot the variation of $\mathtt{SCC}$. We observe that $CC$ has the most disruptive effect on the reduction of $\mathtt{SCC}$ and this happens both in $N_{\mathrm{con}}$ and in $N_{\mathrm{cri}}$.
In Figures \ref{fig:contact-apl-sequential} and \ref{fig:criminal-apl-sequential} we plot the variation of $\mathtt{APL}$ when an increasing fraction $f$ of vertices is neutralized from
$N_{\mathrm{con}}$ and $N_{\mathrm{cri}}$.

From these figures we observe that $CC$ remains the best option to increase $\mathtt{APL}$ in $N_{\mathrm{con}}$; by contrast, the application of $DC$ yields the largest increase in $\mathtt{APL}$ if applied on $N_{\mathrm{cri}}$.

To compare the effectiveness of parallel and sequential police operations, we focus only on the results we achieved on $N_{\mathrm{cri}}$.

From our results, it seems that a sequential police operation has to be preferred to a parallel one: in fact, if we would remove the top $5\%$ mobsters in a sequential police operation we would reduce the size of $\mathtt{SCC}$ up to $65\%$. In contrast, in case of a parallel police operation, we obtain a modest decrease in $\mathtt{SCC}$ of about $4.76\%$
in case of a parallel operation.
Similar results hold for $\mathtt{APL}$.
From this discussion, it seems that sequential strategies should be preferred to parallel ones but, in practice, the identification of the best strategy to dismantle a Mafia gang is a really hard task.
First of all, in fact, police investigation last many years because of the need to collect a large amount of evidence prior to 
arresting individuals. In most cases, law enforcement agencies raid Mafia meetings held to discuss crime strategy and, thus, the end effect of this operation is that some high-caliber mobsters are captured. Therefore, parallel police operations are more realistic (and occurs much more frequently) than sequential ones.
In addition, sequential police operations would achieve the best results if
the Mafia syndicate would slowly react.
In real cases, there are many events which may 
impose a Mafia network to re-organize: on the one hand, in fact, we recount police operations but, on the other hand, we have {\em feuds}, i.e. conflicts between opposite gangs often culminating in the murder of some gang members.

Mafia syndicates, as emerges from judicial documents, are able to {\em instantaneously}  adapt themselves to external events both at the group level (i.e., the Mafia gang may reform their internal organization by electing new bosses) and at the individual level (i.e., criminals may change their behavior or temporarily suspend illicit actions to avoid being targeted by law enforcement agencies).

\section{Conclusions}\label{sec:conclusions}
In this work we presented an experimental analysis of the network structure and resilience of Mafia syndicates. Thanks to collaborations with law enforcement, we were capable of collecting a precious dataset of digital trails and judicial documents that span a period of ten years of investigations of real crimes committed by such syndicates in the North of Sicily. The framework we presented here consists of reconstructing two types of networks, namely a Contact and a Criminal one. The former was constructed from phone-based communications involving suspected individuals, while the latter is based on much stronger evidence of crimes involving actors connected to Mafia syndicates. The sets of actors greatly overlap yet our work highlighted the presence of a small number of high-end criminals who do not appear in the Contact network: this suggests that prominent bosses in Mafia syndicates may not adopt technology to remain off the radars during police investigations. This shows the limits of traditional investigation techniques like tapping, and calls for the adoption of complementary methods that help shed light where data cannot reach. 

Given the unprecedented opportunity to adopt real data for our study, we here focused on investigating the resilience properties of  Contact and Criminal networks. We found that  Criminal networks exhibits and exceptional robustness to targeted attacks,  yet Contact networks are much more vulnerable. We showed that various targeted strategies yield different effects of disruption with different performance, however we provide quantitative evidence that sequential police operation should be preferred to parallel ones, although the latters are way more common and secure, as they expose law enforcement agencies to less risks and potential violent encounters.

Our future research will focus on envisioning strategies of intervention that successfully complement the insights we obtained from this analysis, while from a computational perspective we  aim at defining new methods to identify and predict crimes in the context of Mafia syndicates.

\newpage
\bibliography{bibliography-criminal}

\end{document}